\documentclass[12pt]{article}
\usepackage{amsfonts, latexsym, amscd}
\makeatletter
\@addtoreset{equation}{section}
\makeatother

\def\text#1{\mbox{\rm #1}}

\begin{document}

\title{{{\LARGE {\bf Comparative analysis of the electrogravitational Kepler
problem in GRT and RTG}}}}
\author{{\bf Delia Ionescu}\thanks{Present address: Institute of Mathematics of the Romanian
Academy, P.O. Box 1-764, RO-014700, Bucharest, Romania.
 Email:
Delia.Ionescu@imar.ro}}
\date{}
\maketitle

%\begin{frontmatter}
%\title{Comparative analysis of the electrogravitational Kepler
%problem in GRT and RTG }
%\author{Delia Ionescu\thanksref{now}}
%\thanks[now]{Present address: Zentrum Mathematik, TU M\"{u}nchen, Arcisstr. 21, D-80290 M\"{u}nchen, Germany }
%\ead{ionescu@mathematik.tu-muenchen.de}

%\address{Department of Mathematics, Technical University of Civil Engineering, Bucharest, Romania}

\begin{abstract}
In the framework of Einstein's General Relativity Theory and
of the Relativistic Theory of Gravitation,  the equations governing
the trajectories of charged particles in the  field created by a
 charged mass point are given.
 An analysis
of the shape of the trajectories in both theories is presented. The
first and the second order approximate solutions of
the \textit {electrogravitational Kepler problem} are found in the two theories and the
results are compared with each other.
I have pointed out  the differences between the predictions in the two theories.


\end{abstract}

%\end{frontmatter}
\section{Introduction}

\smallskip

In this paper I study the motion of a charged mass point P having mass $m$
and electric charge $q$, in the electrogravitational field produced by a
charged mass point S having mass $M$ and electric charge $Q$. The \textit{%
electrogravitational Kepler problem} constitutes an analogue to the problem
of the motion of a planet about a fixed sun, under Newtonian attraction.
This problem will be considered, in turn, in the framework of Newton's
Classical Mechanics (CM), of Einstein's General Relativity Theory (GRT) and
of the Relativistic Theory of Gravitation (RTG).

For the sake of comparison with the relativistic versions of the considered
problem, in Section 2 I present  the description of the motion in the
electogravitational field according to CM.

The development of RTG and the differences between this and GRT are
described in detail in [7], [8]. A very important test for a theory of
gravitation is to confirm the astronomical predictions. The predictions of
RTG for the gravitational effects are unique and consistent with the
available experimental data. If the accuracy of the astronomical
measurements could be raised to a level at which the effects of order $\frac{%
velocity^4}{c^4}$ , with $velocity\ll c,$ come into play, it will be
possible to verify the differences between the predictions of the two
theories. Besides, in total contradiction with GRT, static spherically
symmetric bodies in RTG cannot have dimensions less than Schwarzschild
radius. Therefore, the absence of black holes and gravitational collapse in
RTG\ has been confirmed (see, [7], [8], [9] ).

In the framework of GRT, a spherically symmetric solution of the coupled
system of Einstein's Eqs. and Maxwell's Eqs. is that of Nordstr\"{o}m and
Jeffrey (see Wang [14], Section 56). The gravitational radius of the source
point S, as a function depending on $Q^2$ and $M^2,$ has a discontinuity in $%
Q^2=kM^2.$ In Section 3, I present Eqs. of motion of the charged mass point
P in the Nordstr\"{o}m metric. In this section I also present the analysis of the
shape of the orbits in the equatorial plane. What
happens in the vicinity of the gravitational radiu is not presented in detail.
 For this, see  Chandrasekhar [2] and for a complete bibliography of papers on the
geodesics in the Nordstr\"{o}m metric see Sharp [10]. In this section
an approximate  solution of order $\frac{velocity^2}{c^2},$ with
$velocity\ll c$, for the considered problem is given.

The problem of finding the electrogravitational field produced by
the charged mass point S in RTG, was first analyzed by Karabut \& Chugreev [6], but
 assuming only that $kM^2\geq Q^2$. So\'{o}s and I  have reanalyzed the
problem in RTG (see [4]) considering also the possibility $Q^2>kM^2.$ It's important
to analyze this case because the variant is true for the electron. The
analytical form of the solution we found, as well as its domain of
definition, i.e. the gravitational radius $r_g$, depend essentially on the
relation existing between $Q^2$ and $kM^2$. But, in [5] it is shown that this
solution doesn't fulfill the Causality Principle in RTG. Therefore this solution
can't be an acceptable solution for this theory. In [5], I have determined
the unique solution of electrogravitational field produced by a charged mass
point according to RTG. The obtained solution has the same analytical form
for all order relations between $Q^2$ and $kM^2$. The gravitational radius
depends on this relation, but it's a continuous function depending on $Q^2$
and $M^2.$ In Section 4, I present Eqs. of motion of the charged mass point
P in the electrogravitational metric according to RTG. I also analyse  the shape of the orbits in the equatorial plane. I do not analyze
in detail what happens in the vicinity of the gravitational radius. This
problem will be treated in a future paper. In this section, I also write an
approximate solution of order $\frac{velocity^2}{c^2},$ with $velocity\ll
c $, for the considered problem and I compare it with the one obtained in
the framework of GRT in Section 3.

In Section 5, I write solutions to order $\frac{velocity^4}{c^4},$ with $%
velocity\ll c,$ in both theories and then I compare the predictions of the
two theories.

\section{Orbits in the Electrogravitational Field in CM}
To study the problem of motion of the charged mass point P having mass $m$
and electric charge $q$, in the field produced by the charged mass point S
having mass $M$ and electric charge $Q$, we consider a system of coordinates
centered in S. The position of P is denoted by the position vector $\mathbf{r%
}$=$\mathbf{SP}$.

When  P moves in the field produced by S, it is acting on the electrostatic force
$\mathbf{F}_e$ due to $Q$ and the gravitational force $\mathbf{F}_g$ due to $%
M$. The motion of P is governed by Eq.:
\begin{equation}
m\mathbf{a}=\mathbf{F}_{e}+\mathbf{F}_{g},  \label{1}
\end{equation} $\mathbf{a}$ representing the acceleration of P.

The expression of the electrostatic force $\mathbf{F}_{e}$ is given by
Coulomb's law:
\begin{equation}
\mathbf{F}_{e}\frac{qQ}{r^{3}}\mathbf{r} \label{2}
\end{equation}
where $r$ denotes the length of $\mathbf{r}$.
The fact that like charges repel and unlike charges attract each other is
reflected by the direction  of the force $\mathbf{F}_{e}$. When $q,Q$ have the
opposite signs, $\mathbf{F}_{e}$ has the inverse direction  with $\mathbf{r}$.
When they have the same sign, $\mathbf{F}_{e}$ has the same direction  with $%
\mathbf{r.}$

The expression of the gravitational force $\mathbf{F}_{g}$ is given by
Newton's law:
\begin{equation}
\mathbf{F}_{g}=-k\frac{mM}{r^{3}}\mathbf{r}\mbox{\rm ,}  \label{3}
\end{equation} $k$ being the gravitational constant with the empirical value $k$=6,673
$\cdot $10$^{\mbox{\rm -}8}$ gr$^{\mbox{\rm
-}1}$cm$^{3}$s$^{\mbox{\rm -}2}.$ The minus sign in (\ref{3})
indicates that particles attract each other because of the
gravitation.

Introducing (\ref{2}) and (\ref{3}) in (\ref{1}), Eq. of the motion for P
takes the form:
\begin{equation}
m\mathbf{a}=\frac{d\mathbf{v}}{dt}=\frac{d^{2}\mathbf{r}}{dt^{2}}=-k\frac{mM%
}{r^{3}}\mathbf{r}+\frac{qQ}{r^{3}}\mathbf{r}\mbox{\rm ,}  \label{4}
\end{equation} $\mathbf{v}$ representing the velocity of P.
Then
\begin{equation}
\mathbf{r\times }m\mathbf{v}=const=\mathbf{C.}  \label{5}
\end{equation}
Multiplying scalar (\ref{5}) by the vector\textbf{\ }$\mathbf{r}$ we obtain:

\begin{equation}
\mathbf{r\cdot C}=0.  \label{6}
\end{equation}
So the trajectory of P under the action of $\mathbf{F}_e$ and $\mathbf{F}_g$
is situated in a fixed plane which includes S.

We choose the trajectory plane S$xy$. We can describe in this plane the
motion of P using the polar coordinates $r$ and $\theta $ , where $x=r\cos
\theta $, $y=r\sin \theta $. For any position of P in the plane S$xy$, there
is a positive value $r$ and an infinity of values $\theta $ which differ by
an integer multiple of $2\pi $. If P coincides with S, then $r$=0 and $\theta
$ is indefinite.

In the polar coordinates, Eq. (\ref{1}) takes the form:

\begin{equation}
m\left( \frac{d^{2}r}{dt^{2}}-r\left( \frac{d\theta }{dt}\right) ^{2}\right)
=-k\frac{mM}{r^{2}}+\frac{qQ}{r^{2}}  \label{7}
\end{equation}

\begin{equation}
m\left( 2\frac{dr}{dt}\frac{d\theta }{dt}+r\frac{d^{2}\theta }{dt^{2}}%
\right) =0.  \label{8}
\end{equation}

Eq. (\ref{8}) shows that during the motion:

\begin{equation}
r^{2}\frac{d\theta }{dt}=const=J.  \label{9}
\end{equation}

The value of the constant $J,$ which denotes the angular momentum of P per
unit mass, can be determined from the initial conditions. We denote by $r_0$%
, $\theta _0$ the polar coordinates of P at the initial moment $t_0$, by $%
v_0 $ the magnitude of the initial velocity and by $\alpha $ the angle
between $\mathbf{r}_0$ and $\mathbf{v}_0.$ Knowing the expression of the
velocity in polar coordinates we can write:

\begin{equation}
\frac{dr}{dt}(0)=v_{0}\cos \alpha ,\mbox{\rm  }r_{0}\frac{d\theta }{dt}%
(0)=v_{0}\sin \alpha .  \label{10}
\end{equation}

From (\ref{9}) and (\ref{10}), for the constant of the motion $J,$ we get
the value:

\begin{equation}
J=r_{0}v_{0}\sin \alpha .  \label{11}
\end{equation}

As in Kepler's classical problem (see for example [3], Chapter XV), we can
simplify matters by considering $r$ as a function of $\theta $ instead of $t$%
. Any functional relation $r=r(\theta )$ defines a curve in the polar
coordinates system.

Assume $J\neq 0$, so $r$ and $\frac{d\theta }{dt}$ are never 0. Then, the
substitution $u=\frac{1}{r}$ transforms Eq. (\ref{7}) into Binet's
differential Eq. for the orbit of P:

\begin{equation}
\frac{d^{2}u}{d\theta ^{2}}+u=\frac{kM}{J^{2}}-\frac{qQ}{mJ^{2}}.  \label{13}
\end{equation}

This Eq. has the general solution:

\begin{equation}
u=\left( \frac{kM}{J^{2}}-\frac{qQ}{mJ^{2}}\right) +\mathcal{A}\cos (\theta
-\psi ),  \label{18}
\end{equation}
where $\mathcal{A}$ and $\psi $ are constants determined from the initial
conditions. By rotating the coordinates we can make $\psi =0.$

So, if $J\neq 0$, the orbit of P is the conic:

\begin{equation}
\frac{1}{r}=\left( \frac{kM}{J^{2}}-\frac{qQ}{mJ^{2}}\right) +\mathcal{A}%
\cos \theta ,  \label{19}
\end{equation}
S being situated in a focus of this conic.

If the gravitational force $\mathbf{F}_g$ and the electrostatic force $%
\mathbf{F}_e$ have opposite directions and the same magnitude, $kmM-qQ=0,$ than,
from (\ref{19}), P moves along the line:

\begin{equation}
\frac{1}{r}=\mathcal{A}\cos \theta  \label{17}
\end{equation}

Now, we want to see how the nature of the conic (\ref{19}) depends on the
sign of the expression $kmM-qQ$ and on the initial conditions.

Multiplying Eq. (\ref{13}) by $2\frac{du}{d\theta }$ , then integrating,
replacing $u=\frac 1r$ and taking into account (\ref{9}), we obtain the
energy equation:

\begin{equation}
\left( \frac{dr}{dt}\right) ^2=-\frac{J^2}{r^2}+\frac 2m\left( kmM-qQ\right)
\frac 1r+2\mathsf{E},  \label{16}
\end{equation}
$\mathsf{E}$ being the total energy of P per unit mass. This constant of the
motion is also determined from the initial conditions.

Allowing for (\ref{10}), (\ref{11}), we obtain:

\begin{equation}
2\mathsf{E}=v_0^2-\frac 2{mr_0}\left( kmM-qQ\right) .  \label{12}
\end{equation}

Since $\left( \frac{dr}{dt}\right) ^{2}\geq 0$, (\ref{16}) yields $\mathsf{E}
$ $\geq \frac{1}{2}\left( \frac{J^{2}}{r^{2}}-\frac{2}{m}\left(
kmM-qQ\right) \frac{1}{r}\right) .$

The sign of the expression $kmM-qQ$ and the value of $\mathsf{E}$ determine
the range of $r$ and implicitly the shape of the orbit described by P$.$ We
denote the right member of Eq. (\ref{16}) by \textsf{F}$(\frac 1r).$

\textit{Case I)} $\mathsf{E}$ $<0$

This case happens only when $kmM-qQ>0.$ Hence, Eq. \textsf{F}$(\frac
1r)=0 $ has two positive roots, so $r$ oscillates between finite endpoints.
We get  an ellipse as the trajectory of P. If \textsf{F}$(\frac 1r)=0$ has a
double root, then we get a circular orbit. By virtue of (\ref{12}), if:

\[
v_{0}^{2}-\frac{2}{mr_{0}}\left( kmM-qQ\right) <0,%
\mbox{\rm  the orbit is an
ellipse}.
\]

\textit{Case II)} $\mathsf{E}$ $>0$

If $kmM-qQ<0$ then $\mathsf{E}$ $>0,$ but this can also happen if $kmM-qQ>0$%
. Eq. \textsf{F}$(\frac 1r)=0$ has one positive root and one negative root.
Hence, $0<\frac 1r\leq $ a positive root. We obtain a
hyperbola as the trajectory of P. So, by virtue of (\ref{12}), if:

\[
v_{0}^{2}-\frac{2}{mr_{0}}\left( kmM-qQ\right) >0,%
\mbox{\rm  the orbit is a
hyperbola.}
\]

\textit{Case III)} $\mathsf{E}$ $=0$

This case happens only when $kmM-qQ>0.$ Eq. \textsf{F}$(\frac 1r)=0$ has a
root zero and the other root is positive. Hence, $0\leq \frac 1r\leq $
positive root. We obtain a parabola as the trajectory of P. So, by virtue
of (\ref{12}), if:

\[
v_{0}^{2}-\frac{2}{mr_{0}}\left( kmM-qQ\right) =0,%
\mbox{\rm  the orbit is a
parabola}.
\]

In conclusion, if the gravitational force $\mathbf{F}_g$ and the
electrostatic force $\mathbf{F}_e$ have the same direction  or they have opposite
directions, but the magnitude of $\mathbf{F}_g$ is greater then the magnitude of
$\mathbf{F}_e,$ then the orbit described by P in the field produced by S is
an ellipse or a hyperbola or a parabola, all depending on the initial
position and velocity of P. Finally, if $\mathbf{F}_g$ and $\mathbf{F}_e$
have opposite direction s and the magnitude of $\mathbf{F}_g$ is smaller than the
magnitude of $\mathbf{F}_e,$ then P moves on a hyperbola.

\smallskip

\section{Orbits in the Electrogravitational Field in GRT}

\smallskip

Because the basic concepts of Einstein's GRT are so different from those of
Newton's CM, we want to know more about the differences between the
predictions of the two theories in the considered problem.
The study of classical Kepler's problem in GRT is well known (see for example [1], Chapter
VI, Section 3 and [12], Chapter VII, Section 8). In [ 11] , So\'{o}s has
revealed that this problem was one of the main questions taken into account
by Einstein. The capacity of obtaining the correct value for the perihelion
rotation of Mercury has represented a permanent test for the successively
elaborated Einstein's theories of gravitation during the period 1907 -1915.

Let us study the \textit{electrogravitational Kepler problem }in GRT\textit{%
. }The electrogravitational field produced by S, having mass $M$ and
electric charge $Q,$ is described by the following metric (see [14], Section
56):

\begin{eqnarray}
ds^{2}=g_{ij}dx^{i}dx^{j}&=&\left( 1-\frac{2kM}{c^{2}r}+\frac{kQ^{2}}{%
c^{4}r^{2}}\right) \left( dx^{4}\right) ^{2}-\nonumber\\
&&-\frac{1}{1-\frac{2kM}{c^{2}r}+%
\frac{kQ^{2}}{c^{4}r^{2}}}dr^{2}-r^{2}d\varphi ^{2}-r^{2}\sin ^{2}\varphi %
\mbox{\rm  }d\theta ^{2},  \label{1.1}
\end{eqnarray}
$c=$3 $\times $ 10$^{10}$ cm sec$^{\mbox{\rm -}1}$ being the velocity of
light in vacuum.

The metric (\ref{1.1}) is written in the system of coordinates ($x^i)_{i=%
\overline{1,4}}=(r,\varphi ,\theta ,ct)$ centered in S. The coordinates ($%
r,\varphi ,\theta )$ are the spherical coordinates of any point situated in
this field. The domains of definition for these coordinates are: $0\leq
r_g<r<\infty $ , $0< \leq \varphi \leq \pi $ , $0\leq \theta \leq 2\pi $ , $%
-\infty <t<\infty $; $r_g$ representing the gravitational radius of the
point source S. According to GRT, the value of this gravitational radius
depends on the relation between $Q^2$ and $M^2$ in the following manner (see
[14], Section 56):

\begin{equation}
r_{g}=\left\{
\begin{array}{c}
\frac{kM}{c^{2}}+\frac{1}{c^{2}}\sqrt{k^{2}M^{2}-kQ^{2}}\mbox{\rm  , for }%
Q^{2}\leq kM^{2} \\
\mbox{\rm  \qquad \qquad }0\mbox{\rm  \quad \quad , for }Q^{2}>kM^{2}
\end{array}
\right.  \label{1.2}
\end{equation}

The motion of P in the field created by S follows a timelike geodesic line ($%
x^j(s))_{j=\overline{1,4}}=(r(s),\varphi (s),\theta (s),ct(s))$. The
parameter $s$ of this curve is such that $ds^2$ is given by (\ref{1.1}). Eq.
of motion of the charged particle P of mass $m$ and charge $q$, moving in
the field of gravitation ($g_{ij})$ and electromagnetism ($F_{ij}),$ a field
which is not influenced by the particle itself, is (see [13], Section 103):

\begin{equation}
m\left( \frac{d^2x^i}{ds^2}+\Gamma _{jk}^i\frac{dx^j}{ds}\frac{dx^k}{ds}%
\right) +\frac q{c^2}F^i\mbox{\rm  }_j\frac{dx^j}{ds}=0,\mbox{\rm  }i=1,2,3,4
\label{1.3}
\end{equation}

In (\ref{1.3}), $\Gamma _{jk}^i$ are the components of the metric connection
generated by the metric (\ref{1.1}): $\Gamma _{jk}^i=\frac 12g^{il}\left(
\partial _jg_{lk}+\partial _kg_{lj}-\partial _lg_{jk}\right) $ and $F^i$ $%
_j=g^{il}F_{lj}$ are the mixed components of the electromagnetic tensor ($%
F_{ij}).$ For our problem, the nonzero components of the electromagnetic
tensor are (see [14], Section 56):

\begin{equation}
F_{14}=-F_{41}=F^{41}=-F^{14}=\frac{Q}{r^{2}}.  \label{1.4}
\end{equation}

Allowing for (\ref{1.1}), the nonzero components of $F^{i}$ $_{j}$ are:

\begin{equation}
F^4\mbox{\rm  }_1=-f^2\frac Q{r^2},\mbox{\rm  \quad }F^1\mbox{\rm
}_4=-\frac 1{f^2}\frac Q{r^2},  \label{1.5}
\end{equation}
where

\begin{equation}
f^{2}=\frac{1}{1-\frac{2kM}{c^{2}r}+\frac{kQ^{2}}{c^{4}r^{2}}}  \label{1.6}
\end{equation}

Taking into account (\ref{1.1}), the nonzero components $\Gamma
_{jk}^{2},\Gamma _{jk}^{3},$ $\Gamma _{jk}^{4}$ of the metric connection,
which will be used in (\ref{1.3}), are:

\begin{eqnarray}
&&\Gamma _{12}^{2}=\Gamma _{21}^{2}=\Gamma _{13}^{3}=\Gamma _{31}^{3}=\frac{1}{%
r},\mbox{\rm  }\Gamma _{33}^{2}=-\sin \varphi \cos \varphi ,\mbox{\rm  }%
\Gamma _{23}^{3}=\Gamma _{32}^{3}=\cot \varphi ,\nonumber\\
&&\mbox{\rm  }\Gamma
_{14}^{4}=\Gamma _{41}^{4}=-\frac{1}{f}\frac{df}{dr}  \label{1.7}
\end{eqnarray}

Thus, allowing for (\ref{1.7}) and setting $i=2$ in (\ref{1.3}), we get:

\begin{equation}
\frac{d^{2}\varphi }{ds^{2}}+\frac{2}{r}\frac{d\varphi }{ds}\frac{dr}{ds}%
-\sin \varphi \cos \varphi \left( \frac{d\theta }{ds}\right) ^{2}=0.
\label{1.8}
\end{equation}

By an appropriate orientation of the axes, we can initially have $\varphi
(s_0)=\frac \pi 2$ and $\frac{d\varphi }{ds}(s_0)=0$. Thus the solution of
Eq. (\ref{1.8}) is:

\begin{equation}
\varphi (s)=\frac{\pi }{2}  \label{1.9}
\end{equation}

So we can see that as in the classical case, the orbit lies in a plane.

Considering $i=3$ in (\ref{1.3}) and taking into account (\ref{1.7}), (\ref
{1.9}) we obtain:

\begin{equation}
\frac{d^{2}\theta }{ds^{2}}+\frac{2}{r}\frac{dr}{ds}\frac{d\theta }{ds}=0.
\label{1.10}
\end{equation}

Integrating this Eq., we find:

\begin{equation}
r^2\frac{d\theta }{ds}=const=L.  \label{1.11}
\end{equation}
This Eq. is similar to Eq. (\ref{9}), hence we can call $L$ the angular
momentum of P per unit mass.

We set $i=4$ in (\ref{1.3}) and from (\ref{1.5}), (\ref{1.7}), we get:

\begin{equation}
\frac{d^{2}x^{4}}{ds^{2}}-2\frac{1}{f}\frac{df}{ds}\frac{dx^{4}}{ds}=f^{2}%
\frac{qQ}{mc^{2}r^{2}}\frac{dr}{ds}.  \label{1.12}
\end{equation}

Eq. (\ref{1.12}) integrates to:

\begin{equation}
\frac{dx^{4}}{ds}=\left( E-\frac{qQ}{mc^{2}}\frac{1}{r}\right) f^{2},
\label{1.13}
\end{equation}
$E$ being a constant.

To obtain $\frac{dr}{ds}$ we can consider $i=1$ in (\ref{1.3}) but it is
more convenient to divide the line element (\ref{1.1}) by $ds^2$. Allowing
for (\ref{1.9}), (\ref{1.11}), (\ref{1.13}), we find Eq.:

\begin{equation}
f^{2}\left( \frac{dr}{ds}\right) ^{2}+\frac{L^{2}}{r^{2}}-f^{2}\left( E-%
\frac{qQ}{mc^{2}}\frac{1}{r}\right) ^{2}+1=0,  \label{1.14}
\end{equation}
which is analogous to the classical energy Eq. (\ref{16}).

As in the problem considered in the framework of CM, we'll consider $r$ as a
function of $\theta $ instead of $s$. Thus, taking into account (\ref{1.11}%
), we have:

\begin{equation}
\frac{dr}{ds}=\frac{dr}{d\theta }\frac{d\theta }{ds}=\frac{L}{r^{2}}\frac{dr%
}{d\theta }.  \label{1.15}
\end{equation}

Putting

\begin{equation}
u=\frac 1r  \label{1.16}
\end{equation}
and considering the case when $L\neq 0$, Eq. (\ref{1.14}) becomes:

\begin{eqnarray}
\left( \frac{du}{d\theta }\right) ^{2} &=&-\frac{kQ^{2}}{c^{4}}u^{4}+\frac{%
2kM}{c^{2}}u^{3}-\left[ 1+\frac{Q^{2}}{L^{2}m^{2}c^{4}}\left(
km^{2}-q^{2}\right) \right] u^{2}+  \nonumber \\
&&+\left( \frac{2kM}{L^{2}c^{2}}-E\frac{2qQ}{mL^{2}c^{2}}\right) u-\frac{%
1-E^{2}}{L^{2}}  \label{1.17}
\end{eqnarray}

Eq. (\ref{1.17}) governs the geometry of the orbits described by P in the
plane $\varphi =\frac{\pi }{2}.$

We denote the right member of Eq. (\ref{1.17}) by $F(u).$ It is also clear
that the disposition of the roots of Eq. $F(u)=0$ will determine the shape
of the orbit. By (\ref{1.17}), $F\left( u\right) \geq 0$ throughout the
orbit and $F\left( u\right) $ tends to $-\infty $ for very large values of $%
u $. It follows that $F\left( u\right) =0$ has four real roots or two real
roots and a complex conjugate pair. A root signifies a turning point where $%
\frac{du}{d\theta }$ changes the sign. A negative root has no physical
meaning, so  at least one positive root should occur.

The consideration of Eq. (\ref{1.17}) is conveniently separated into the
following parts: $E^{2}<1,E^{2}>1,E^{2}=1.$ These distinctions determine
whether the orbits are bound or unbound (i. e. whether along the orbit $r$
remains bounded or not). These classes of orbits are characterized by total
energies (exclusive of the rest energy) which are negative, positive or
zero.

We denote $u_{1},u_{2},u_{3},u_{4}$ the zeros of $F\left( u\right) ,$ with $%
u_{1}<u_{2}<u_{3}<u_{4}$ if they are all real. We have:

\begin{eqnarray}
u_{1}+u_{2}+u_{3}+u_{4} &=&\frac{2Mc^{2}}{Q^{2}}>0  \label{v1} \\
u_{1}u_{2}u_{3}u_{4} &=&\frac{1-E^{2}}{L^{2}}\frac{c^{4}}{kQ^{2}}  \label{v2}
\end{eqnarray}

\textit{Case I)} $E^{2}<1$

In this case, from (\ref{v1}), (\ref{v2}), if there are only two real roots,
they must be positive. If all the roots are real, then two of them or all
four must be positive. We also have $F(u)<0$ for $u$ tends to $0.$ Thus, we
get an orbit of elliptic type with $u$ oscillating in the range $u_1\leq
u\leq u_2$\ or $u_3\leq u\leq u_4$ ($u_1$ or $u_3$, corresponds to aphelion,
$u_2$ or $u_4$ corresponds to perihelion).

\textit{Case II) }$E^{2}>1$

In this case, by virtue of (\ref{v1}), (\ref{v2}), one (for example $u_4)$
or three roots (for example $u_2,u_3,u_4)$ must be positive. We also have $%
F(u)>0$ for $u$ tends to $0$. Thus, we get an orbit of hyperbolic type
restricted to the interval $0<u\leq u_4$ or $0<u\leq u_2$ or an orbit of
elliptic type restricted to the interval $u_3\leq u\leq u_4$.

\textit{Case III)} $E^{2}=1$

In this case, one of the solutions is zero. Thus, we get an orbit of parabolic
type restricted to the interval $0\leq u\leq u_2$ or an orbit
of elliptic type restricted to the interval $u_3\leq u\leq u_4.$

In the special case of double roots, we get a circular orbit.

From (\ref{1.2}), if $Q^2>kM^2$, then $r_g=0$. The biggest positive
solution of the equation $F(u)=0$ can take any large value and the
non-capture orbits occur. But if $Q^2\leq kM^2$, then from (\ref{1.2}), $%
r_g= $ $\frac{kM}{c^2}+\frac 1{c^2}\sqrt{k^2M^2-kQ^2}$, so it will be
possible that the biggest positive solution of $F(u)=0$ overpasses $\frac
1{r_g}.$ From the viewpoint of GRT (for a complete bibliography of papers on
the geodesics in the Nordstr\"{o}m metric (\ref{1.1}) see Sharp [10]; see
also Chandrasekhar [2], Chapter 5),  the particle will cross the
horizon $r=r_g$ in this case only in the inside direction and its trajectory will
formally terminate at this turning point. In addition, from (\ref{1.14}), at
this turning point $1-\frac{2kM}{c^2r}+\frac{kQ^2}{c^4r^2}$ must be
positive, so this turning point is in the interval $(0,$ $\frac{kM}{c^2}%
-\frac 1{c^2}\sqrt{k^2M^2-kQ^2}).$ For an external observer, P will take an
infinite time to reach the horizon $r=r_g$, but the falling observer with P
will cross the horizon $r=r_g$ and reach the turning point in a finite time
which is its own proper time$.$

Let us now explore Eq. (\ref{1.17}) with the view to find a solution.
 The exact solution of this Eq. expresses the angle $\theta $ as an
elliptic integral of $u=\frac 1r$ and conversely it gives $u$ as an implicit
function of $\theta .$ Unfortunately, this implicit form of the solution
doesn't make evident the approximate classical form of the trajectory. To
establish a closer connection with the classical problem, we differentiate
Eq. (\ref{1.17}) with respect to $\theta .$ One possible solution is
obtained by setting the common factor $\frac{du}{d\theta }$ equal to zero.
This yields $u=const,$ so $r=const.$ Thus the circular motion occurs also in
GRT. Removing the common factor 2$\frac{du}{d\theta }$, we obtain:

\begin{equation}
\hspace{-1cm}\frac{d^{2}u}{d\theta ^{2}}+u=-\frac{2kQ^{2}}{c^{4}}u^{3}+\frac{3kM}{c^{2}}%
u^{2}+\frac{Q^{2}}{L^{2}c^{4}m^{2}}\left( q^{2}-km^{2}\right) u+\frac{kM}{%
L^{2}c^{2}}-E\frac{qQ}{mL^{2}c^{2}}.  \label{1.18}
\end{equation}

In the case of slow motion in weak gravitational fields, Eq.(\ref{1.18})
must reduce to the classical Eq. (\ref{13}). Indeed, for a slowly moving
particle in a weak field, we have $\frac{dx^4}{ds}$ $\simeq 1$ and $E\simeq 1+\frac{%
\mathsf{E}}{c^2},$ where $\mathsf{E}$ is the total energy of P per unit
mass, given by (\ref{16}). Thus, from (\ref{9}) and (\ref{1.11}), we get:

\begin{equation}
\frac 1{L^2c^2}=\frac 1{r^2\left( \frac{d\theta }{ds}\right) ^2c^2}=\frac
1{r^2\left( \frac{d\theta }{dt}\right) ^2\left( \frac{dt}{ds}\right)
^2c^2}=\frac 1{J^2\left( \frac{dx^4}{ds}\right) ^2}\simeq \frac 1{J^2}.
\label{1.20}
\end{equation}
and taking $E\simeq 1$ , Eq. (\ref{1.18}) reduces to Eq. (\ref{13}) in the
case of slow motion in weak gravitational fields.

We notice that relativistic Eq. (\ref{1.18}) differs from the classical Eq. (%
\ref{13}) through the addition of three terms containing $u$, and it has a
slightly different constant term.

For the sake of simplicity, in the view of (\ref{1.20}) and for $E\simeq 1,$
let us now investigate the orbital Eq.:

\begin{equation}
\hspace{-0.5cm}\frac{d^{2}u}{d\theta ^{2}}+u=\frac{kM}{J^{2}}-\frac{qQ}{mJ^{2}}+\frac{Q^{2}%
}{J^{2}c^{2}m^{2}}\left( q^{2}-km^{2}\right) u+\frac{3kM}{c^{2}}u^{2}-\frac{%
2kQ^{2}}{c^{4}}u^{3}.  \label{1.21}
\end{equation}

Let us evaluate the order of magnitude of the three terms containing $u$
from the right side of Eq. (\ref{1.21}). The order of magnitude of $u$ is $%
\frac 1l,$ where $l$ is a length. The order of magnitude of $kMu$ is $v_1^2,$
$v_1$ being considered a velocity much smaller than the velocity of light in
vacuum. Thus the term of second order in $u$ has the magnitude $\frac{v_1^2}{%
c^2}\frac 1l.$ The order of magnitude of $\frac{Q^2}{J^2m^2}\left(
q^2-km^2\right) $ is also the square of a velocity $v_2,$ considered much
smaller than the velocity of light in vacuum. Thus, the term of the first order
in $u$ has the order of magnitude $\frac{v_2^2}{c^2}\frac 1l.$ Finally, $%
kQ^2u^2$ has the order of magnitude $v_3^4,$where $v_3$ is considered a
velocity much smaller than the velocity of light in vacuum. So, the term of the
third order in $u$ has the order of magnitude $\frac{v_3^4}{c^4}\frac 1l.$
The term of order $\frac{v_3^4}{c^4}$ is very small compared with the
unity, so, if we want to find a solution of Eq. (\ref{1.21}) to order $\frac{%
velocity^2}{c^2}$, $velocity\ll c$, we neglect the last term in Eq. (\ref
{1.21}). We define the small dimensionless quantities:

\begin{equation}
\varepsilon =\frac{Q^{2}}{J^{2}m^{2}}\left( q^{2}-km^{2}\right) \frac{1}{%
c^{2}},\mbox{\rm  \quad }\delta =\frac{k^{2}M^{2}}{J^{2}c^{2}}  \label{1.22}
\end{equation}

Thus, Eq. (\ref{1.21}) becomes:

\begin{equation}
\frac{d^{2}u}{d\theta ^{2}}+u\left( 1-\varepsilon \right) =\frac{kM}{J^{2}}-%
\frac{qQ}{mJ^{2}}+3\delta \frac{J^{2}}{kM}u^{2}.  \label{1.23}
\end{equation}

Let us find an approximate  solution of this nonlinear Eq., by a
perturbation approach.

To solve this, we assume a solution of the form:

\begin{equation}
u(\theta )=u_{o}(\theta )+\varepsilon V(\theta )+\delta W(\theta
)+O(\varepsilon ^{2})+O(\delta ^{2})+O(\varepsilon \delta ).  \label{1.24}
\end{equation}

Substituting this form for $u$ in the differential Eq.(\ref{1.23}) and
keeping only the terms of order 0 and 1 in $\varepsilon $ and $\delta $, we
find:

\begin{equation}
\hspace{-0.9cm}\frac{d^{2}u_{o}}{d\theta ^{2}}+\varepsilon \frac{d^{2}V}{d\theta ^{2}}%
+\delta \frac{d^{2}W}{d\theta ^{2}}+u_{o}+\varepsilon V+\delta W-\varepsilon
u_{o}=\frac{kM}{J^{2}}-\frac{qQ}{mJ^{2}}+3\delta \frac{J^{2}}{kM}u_{o}^{2}.
\label{1.25}
\end{equation}
Equating the zeroth order terms in $\varepsilon $ and $\delta $ we get:

\begin{equation}
\frac{d^{2}u_{o}}{d\theta ^{2}}+u_{o}=\frac{kM}{J^{2}}-\frac{qQ}{mJ^{2}}.
\label{1.26}
\end{equation}
which is the classical Eq. (\ref{13}). The solution of this Eq. is given by (%
\ref{19}).

Equating the first order terms in $\varepsilon $ and taking into account (%
\ref{19}), we get:

\begin{equation}
\frac{d^{2}V}{d\theta ^{2}}+V=\frac{kM}{J^{2}}-\frac{qQ}{mJ^{2}}+\mathcal{A}%
\cos \theta .  \label{1.28}
\end{equation}

We need only a nonhomogeneous solution to this Eq., since the zeroth order
solution already contains a term $\mathcal{A}$cos$\theta $, which is the
general solution to the homogeneous Eq.. Thus, we find:

\begin{equation}
V(\theta )=V_{1}(\theta )+V_{2}(\theta ),  \label{1.29}
\end{equation}
where

\begin{eqnarray}
V_{1}(\theta ) =\frac{kM}{J^{2}}-\frac{qQ}{mJ^{2}},\mbox{\rm  } \qquad
V_{2}(\theta ) =\frac{\mathcal{A}}{2}\theta \sin \theta .  \label{1.30}
\end{eqnarray}

Similarly, equating the first order terms in $\delta $ and by virtue of (\ref
{19}), we obtain:

\begin{eqnarray}
\frac{d^{2}W}{d\theta ^{2}}+W &=&3\frac{J^{2}}{kM}\left( \frac{kM}{J^{2}}-%
\frac{qQ}{mJ^{2}}\right) ^{2}+\frac{3}{2}\frac{J^{2}}{kM}\mathcal{A}^{2} + \nonumber \\
&&+6
\frac{J^{2}}{kM}\left( \frac{kM}{J^{2}}-\frac{qQ}{mJ^{2}}\right) \mathcal{A}%
\cos \theta +\frac{3}{2}\frac{J^{2}}{kM}\mathcal{A}^{2}\cos 2\theta .  \label{1.31}
\end{eqnarray}
with the nonhomogeneous solution:

\begin{equation}
W(\theta )=W_{1}(\theta )+W_{2}(\theta )+W_{3}(\theta ),  \label{1.32}
\end{equation}
where
\begin{eqnarray}
W_{1}(\theta ) &=&3\frac{J^{2}}{kM}\left( \frac{kM}{J^{2}}-\frac{qQ}{mJ^{2}}%
\right) ^{2}+\frac{3}{2}\frac{J^{2}}{kM}\mathcal{A}^{2}  \nonumber \\
W_{2}(\theta ) &=&3\frac{J^{2}}{kM}\left( \frac{kM}{J^{2}}-\frac{qQ}{mJ^{2}}%
\right) \mathcal{A}\theta \sin \theta  \nonumber \\
W_{3}(\theta ) &=&-\frac{1}{2}\frac{J^{2}}{kM}\mathcal{A}^{2}\cos 2\theta .
\label{1.33}
\end{eqnarray}

Introducing (\ref{19}), (\ref{1.29}), (\ref{1.32}) into (\ref{1.24}), we
find the solution for the orbit to first order in $\varepsilon $ and $\delta
$:

\begin{eqnarray}
\hspace{-0.9cm}u(\theta ) &=&\left[ \frac{kM}{J^2}-\frac{qQ}{mJ^2}+\varepsilon \left( \frac{%
kM}{J^2}-\frac{qQ}{mJ^2}\right) +\delta \frac{3J^2}{kM}\left( \left( \frac{kM%
}{J^2}-\frac{qQ}{mJ^2}\right) ^2+\frac{\mathcal{A}^2}2\right) \right] +
\nonumber \\
\hspace{-1cm}+&&\left[ \mathcal{A}\cos \theta -\delta \frac{J^2\mathcal{A}^2}{2kM}\cos
2\theta \right] +\left[ \frac \varepsilon 2+\delta \frac{3J^2}{kM}\left(
\frac{kM}{J^2}-\frac{qQ}{mJ^2}\right) \right] \mathcal{A}\theta \sin \theta .
\label{1.34}
\end{eqnarray}

In the solution (\ref{1.34}), only the last term is nonperiodic. To clarify
further the effect of this nonperiodic term, we note that, to the first order in
$\varepsilon $ and $\delta $,

\begin{eqnarray}
\cos \left( \theta -\left( \frac \varepsilon 2
+\delta \frac{3J^2}{kM}\left(
\frac{kM}{J^2}\right.\right.\right.&-&\left.\left.\left.\frac{qQ}{mJ^2}\right) \right) \theta \right)=\cos \theta+\nonumber\\
&+&\left( \frac \varepsilon 2+\delta \frac{3J^2}{kM}\left( \frac{kM}{J^2}-%
\frac{qQ}{mJ^2}\right) \right) \theta \sin \theta ,  \label{1.35}
\end{eqnarray}
so the solution may be written as:

\begin{eqnarray}
u(\theta )&=&\frac{kM}{J^{2}}-\frac{qQ}{mJ^{2}}+\mathcal{A}\cos \left( \theta
-\left( \frac{\varepsilon }{2}+\delta \frac{3J^{2}}{kM}\left( \frac{kM}{J^{2}%
}-\frac{qQ}{mJ^{2}}\right) \right) \theta \right) +\nonumber\\
&+& (%
\mbox{\rm periodic terms of
order }\varepsilon \ \mbox{\rm   and }\delta )  \label{1.36}
\end{eqnarray}
The effect of the last term is to introduce small periodic variations in the
radial distance of P.

In the case of an orbit of elliptic type, the effect of the second term can
be clarified. The small differences between relativistic orbit and the
Newtonian ellipse $\frac{kM}{J^2}-\frac{qQ}{mJ^2}+\mathcal{A}\cos \theta $
are due to this term which influences the perihelion position of P.
The perihelion of P is the point of closest approach to S. This occurs when $%
u$ is maximum. From (\ref{1.36}) we see that $u$ is maximum when:

\begin{equation}
\theta \left( 1-\left( \frac{\varepsilon }{2}+\delta \frac{3J^{2}}{kM}\left(
\frac{kM}{J^{2}}-\frac{qQ}{mJ^{2}}\right) \right) \right) =2\pi n.
\label{1.37}
\end{equation}

Keeping the terms to the first order in $\varepsilon $ and $\delta ,$ the
interval between successive perihelia is:

\begin{equation}
\triangle \theta =2\pi \left( 1+\frac{\varepsilon }{2}+\delta \frac{3J^{2}}{%
kM}\left( \frac{kM}{J^{2}}-\frac{qQ}{mJ^{2}}\right) \right) ,  \label{1.38}
\end{equation}
instead of $2\pi $ like in periodic motion. So, according to the notations (%
\ref{1.22}), the shift of the perihelion is approximately:

\begin{equation}
\triangle \theta -2\pi =6\pi \frac{k^{2}M^{2}}{c^{2}J^{2}}-6\pi \frac{qQkM}{%
mJ^{2}c^{2}}+\pi \frac{Q^{2}\left( q^{2}-km^{2}\right) }{J^{2}m^{2}c^{2}}.
\label{1.39}
\end{equation}

Therefore, the orbit of P is to be regarded as an ellipse which rotates
slowly. We see that if $Q$, $q$ are zero, we find the known formula for the
advance of perihelion per revolution, which is one of the famous formulas of GRT. From
(\ref{1.39}), we see that when the charges of P and S are taken into
account, if
\begin{equation}
\mbox{\rm   }6kmM\left( kmM-qQ\right) +Q^2\left( q^2-km^2\right) >0
\label{1.40}
\end{equation}
then the ellipse rotates in the direction  in which it is described and if

\begin{equation}
6kmM\left( kmM-qQ\right) +Q^2\left( q^2-km^2\right) <0  \label{1.41}
\end{equation}
then the ellipse rotates in the opposite direction  in which it is described.

\section{Orbits in the Electrogravitational Field in RTG}

GRT encounters serious difficulties with the evaluation of the physical
characteristics of the gravitational field and the formulation of the
energy-momentum conservation laws. Combining Poincar\'{e}'s idea of the
gravitational field as a Faraday-Maxwell physical field with Einstein's idea
of a Riemannian space-time geometry, Logunov and his co-workers have
elaborated a \textit{new }relativistic theory of gravitation, named RTG (see
[ 7], [8 ]), in the framework of the Special Theory of Relativity (SRT). In
this theory, the Minkowski space-time is a fundamental space that
incorporates all physical fields, including gravitation. The gravitational
field is described by a second order symmetric tensor $\phi ^{ij}$,
possessing energy-momentum density, rest mass and polarization states
corresponding to spin 2 and 0. Owing to the action of this field, an
effective Riemannian space-time $g_{ij}$ arises. GRT characterizes the
gravitational field by the metric tensor $g_{ij},$ whereas in RTG it is
determined by the tensor value $\phi ^{ij},$ the effective Riemannian
space-time being constructed with the help of the field $\phi ^{ij}$ and of
the Minkowski metric tensor to fix the choice of the coordinate system. The
construction rule is the following: $\tilde{g}^{ij}=\sqrt{-g}g^{ij}=\sqrt{%
-\gamma }\gamma ^{ij}+\sqrt{-\gamma }\phi ^{ij},\mbox{\rm  }g=\det
(g_{ij}),\gamma =\det (\gamma _{ij}).$ Metric properties are determined by
the effective Riemannian space-time tensor $g_{ij}$ in the presence of a
gravitational field and by the Minkowski space-time tensor $\gamma _{ij}$ in
the absence of this field. The interaction between tensor gravitational
field and matter can be introduced as though it deformed the Minkowski
space, changing the metric properties without affecting the causality. The
Causality Principle in RTG affirms that the light cone in the effective
Riemannian space-time does not go beyond the causality cone of the
Minkowskian space-time. For the differential laws of RTG and the analytic
formulation of the Causality Principle in RTG see [7], [8].

The problem of finding the field of an electrically charged mass point
having mass $M$ and electric charge $Q,$ was first analyzed by Karabut \&
Chugreev in [6], but only assuming $kM^2\geq Q^2$. So\'{o}s and I
have reanalyzed this problem in RTG  in [4], considering also the possibility $%
Q^2>kM^2.$ It is important to analyze this case because the variant is true
for the electron. The analytical form of the solution we found, as well as
its domain of definition, i.e. the gravitational radius $r_g$, depend
essentially on the relation existing between $Q^2$ and $kM^2$. But in [5] I
have shown that the solution obtained by us doesn't fulfill the Causality
Principle in RTG. So, this solution can not be an acceptable solution in this
theory. I have determined in [5] the unique solution of electrogravitational
field produced by a charged mass point according to RTG. The obtained
solution has the same analytical form for all order relations between $Q^2$
and $kM^2$. The gravitational radius depends on this relation but it is a
continuous function depending on $Q^2$ and $kM^2.$

Solving the coupled system of RTG's Eqs. and Maxwell's Eqs, and taking into
account the Causality Principle in RTG, we get the following effective
Riemannian space-time due to the elecrogravitational field produced by a
charged mass point with mass $M$ and charge $Q:$

\begin{eqnarray}
&&ds^{2} =g_{ij}dx^idx^j=\left( 1-\frac{2kM}{c^{2}\left( r+\frac{kM}{%
c^{2}}\right) }+\frac{kQ^{2}}{c^{4}\left( r+\frac{kM}{c^{2}}\right) ^{2}}%
\right) \left( dx^{4}\right) ^{2} -\nonumber \\
&&-\frac{1}{1-\frac{2kM}{c^{2}\left( r+\frac{%
kM}{c^{2}}\right) }+\frac{kQ^{2}}{c^{4}\left( r+\frac{kM}{c^{2}}\right) ^{2}}%
}dr^{2}-\left( r+\frac{kM}{c^{2}}\right) ^{2}\left( d\varphi ^{2}+\sin
^{2}\varphi d\theta ^{2}\right) .  \label{2.1}
\end{eqnarray}

The metric (\ref{2.1}) is written in the Minkowskian system of coordinates ($%
x^i)_{i=\overline{1,4}}=(r,\varphi ,\theta ,ct)$ centered in S. The domains
of definition for these coordinates are: $0\leq r_g<r<\infty $ , $0<\leq
\varphi \leq \pi $ , $0\leq \theta \leq 2\pi $ , $-\infty <t<\infty $; $r_g$
representing the gravitational radius of the point source S. According to
RTG, the value of this gravitational radius depends on the relation between $%
Q^2$ and $kM^2$ in the following manner (see [5]):

\begin{equation}
r_{g}=\left\{
\begin{array}{c}
\frac{1}{c^{2}}\sqrt{k^{2}M^{2}-kQ^{2}}\mbox{\rm  , for }Q^{2}\leq kM^{2} \\
\mbox{\rm  \qquad \qquad }0\mbox{\rm  \quad \quad , for }kM^{2}<Q^{2}<2kM^{2}
\\
\frac{1}{2c^{2}M}\left( Q^{2}-2kM^{2}\right) \mbox{\rm  , for }Q^{2}\geq
2kM^{2}
\end{array}
\right.  \label{2.2}
\end{equation}

We notice in (\ref{2.2}), that the function $r_g$ depending on $Q^2$ and $%
kM^2$ is a continuous one, which is not the case for GRT's function $r_g$
from (\ref{1.2}) .

The metric of the Minkowski space-time in which we happen to be when the
gravitational field is switched off is:

\begin{equation}
d\sigma ^2=\gamma _{ij}dx^idx^j=c^2dt^2-dr^2-r^2d\theta ^2-r^2\sin ^2\theta
d\varphi ^2.  \label{*}
\end{equation}

According to the principles of RTG, the motion of matter under the action of
a gravitational field in the Minkowski space-time is identical to its motion
in the effective Riemannian space-time with the metric $g_{ij}$. Thus, the
equation of motion of the charged particle P of mass $m$ and charge $q$,
moving in the field produced by S, is:

\begin{equation}
m\left( \frac{d^2x^i}{ds^2}+\Gamma _{jk}^i\frac{dx^j}{ds}\frac{dx^k}{ds}%
\right) +\frac q{c^2}F^i\mbox{\rm  }_j\frac{dx^j}{ds}=0,\mbox{\rm  }i=1,2,3,4
\label{2.3}
\end{equation}

Eqs.(\ref{2.3}) are similar to Einstein's Eqs.(\ref{1.3}), the one important
difference being that in RTG all field variables depend on the universal
spatial-temporal coordinates in the Minkowski space-time. $\Gamma _{jk}^i$
are the components of the metric connection generated by the effective
Riemannian metric (\ref{2.1}) and $F^i$ $_j=g^{il}F_{lj}$ are the mixed
components of the electromagnetic tensor ($F_{ij}).$ For our problem, the
nonzero components of the electromagnetic tensor are (see [5]):

\begin{equation}
F_{14}=-F_{41}=F^{41}=-F^{14}=\frac{Q}{\left( r+\frac{kM}{c^{2}}\right) ^{2}}%
.  \label{2.4}
\end{equation}

Allowing for (\ref{2.1}), the nonzero components of $F^{i}$ $_{j}$ are:

\begin{equation}
F^4\mbox{\rm  }_1=-\frak{f}^2\frac Q{\left( r+\frac{kM}{c^2}\right) ^2},%
\mbox{\rm  \quad }F^1\mbox{\rm  }_4=-\frac 1{\frak{f}^2}\frac Q{\left( r+%
\frac{kM}{c^2}\right) ^2},  \label{2.5}
\end{equation}
where

\begin{equation}
\frak{f}^{2}=\frac{1}{1-\frac{2kM}{c^{2}\left( r+\frac{kM}{c^{2}}\right) }+%
\frac{kQ^{2}}{c^{4}\left( r+\frac{kM}{c^{2}}\right) ^{2}}}  \label{2.6}
\end{equation}

Taking into account (\ref{2.1}), the nonzero components $\Gamma
_{jk}^{2},\Gamma _{jk}^{3},$ $\Gamma _{jk}^{4}$ of the metric connection,
which will be used in (\ref{2.3}), are:

\begin{eqnarray}
\Gamma _{12}^{2} &=&\Gamma _{21}^{2}=\Gamma _{13}^{3}=\Gamma _{31}^{3}=\frac{%
1}{\left( r+\frac{kM}{c^{2}}\right) },\ \Gamma _{33}^{2}=-\sin
\varphi \cos \varphi ,\mbox{\rm  },\mbox{\rm  }  \nonumber \\
\Gamma _{23}^{3}&=& \Gamma _{32}^{3}=\cot
\varphi,\ \Gamma _{14}^{4} =\Gamma _{41}^{4}=-\frac{1}{\frak{f}}\frac{d\frak{f}}{dr}
\label{2.7}
\end{eqnarray}

Using the same procedures as in Section 3, we find:

\begin{equation}
\varphi (s)=\frac{\pi }{2}  \label{2.8}
\end{equation}

So we can see that like in CM and GRT, the orbit lies in a plane.

Integrating Eq. (\ref{2.3}) for $i=3$ and taking into account (\ref{2.7}), (%
\ref{2.8}) we obtain:

\begin{equation}
\left( r+\frac{kM}{c^2}\right) ^2\frac{d\theta }{ds}=const=\mathcal{L}.
\label{2.9}
\end{equation}

Allowing for (\ref{2.5}), (\ref{2.7}), Eq. (\ref{2.3}) for $i=4$ integrates
to

\begin{equation}
\frac{dx^4}{ds}=\left( \frak{E}\mbox{\rm  }-\frac{qQ}{mc^2}\frac 1{\left( r+%
\frac{kM}{c^2}\right) }\right) \frak{f}^2  \label{2.10}
\end{equation}
$\frak{E}$ being a constant.

Dividing the line element (\ref{2.1}) by $ds^2$ and allowing for (\ref{2.8}%
), (\ref{2.9}), (\ref{2.10}) , we find the following Eq., which is analogous
to the classical energy Eq. (\ref{16}), according to RTG.:

\begin{equation}
\frak{f}^2\left( \frac{dr}{ds}\right) ^2+\frac{\mathcal{L}^2}{\left( r+\frac{%
kM}{c^2}\right) ^2}-\frak{f}^2\left( \frak{E}\mathcal{-}\frac{qQ}{mc^2}\frac
1{\left( r+\frac{kM}{c^2}\right) }\right) ^2+1=0  \label{2.11}
\end{equation}

As in the problem considered in the framework of CM or GRT, we'll consider $%
r $ as a function of $\theta $ instead of $s$. Thus, taking into account (%
\ref{2.9}):

\begin{equation}
\frac{dr}{ds}=\frac{dr}{d\theta }\frac{d\theta }{ds}=\frac{\mathcal{L}}{%
\left( r+\frac{kM}{c^2}\right) ^2}\frac{dr}{d\theta }.  \label{2.12}
\end{equation}
and putting

\begin{equation}
Z=\frac 1{r+\frac{kM}{c^2}},  \label{2.13}
\end{equation}
for the case when $\mathcal{L}\neq 0,$ Eq. (\ref{2.11}) becomes:

\begin{eqnarray}
\left( \frac{dZ}{d\theta }\right) ^2 &=&-\frac{kQ^2}{c^4}Z^4+\frac{2kM}{c^2}%
Z^3-\left[ 1+\frac{Q^2}{\mathcal{L}^2m^2c^4}\left( km^2-q^2\right) \right]
Z^2+  \nonumber \\
&&+\left( \frac{2kM}{\mathcal{L}^2c^2}-\frak{E}\frac{2qQ}{m\mathcal{L}^2c^2}%
\right) Z-\frac{1-\frak{E}^2}{\mathcal{L}^2}  \label{2.14}
\end{eqnarray}

Eq. (\ref{2.14}) governs, according to RTG, the geometry of the orbits
described by P in the plane $\varphi =\frac \pi 2$. As in GRT, denoting by $%
\mathcal{F}(Z)$ the right member of this Eq., the range for $r$ and the
shape of the orbit are determined by the disposition of the roots of Eq. $%
\mathcal{F}(Z)=0$ and by the value of $\mathcal{\ }\frak{E}^2.$ Eq. $%
\mathcal{F}(Z)=0$ has the same roots as Eq. $F(u)=0$ from (\ref{1.17}), so
the discussion concerning the disposition of these roots depending on the
value of $\frak{E}^2,$ rests the same as in Section 3. Thus, if $\frak{E}%
^2<1 $ we obtain elliptic type orbits, if $\frak{E}^2>1$ we get hyperbolic
or elliptic type orbits and if $\frak{E}^2=1$ we get parabolic or elliptic
type orbits. For these orbits, $Z$ oscillates in the same range as $u,$ so
the range of $r$ is moved back with $\frac{kM}{c^2}.$

The substantial difference between the solution in GRT and RTG is
established in the region close to the gravitational radius $r_g,$ given by (%
\ref{2.2})$.$ This case must be treated as Logunov and Mestvirishvili have
made in [9] for the case of Schwarzschild metric. For this purpose, we must
consider the solution of the system of RTG's Eqs. with a nonzero graviton
mass and Maxwell's Eqs. All details concerning this problem will be treated
in a future paper. For now, the following remark can be made here: The
particle can not continue its trajectory beyond the horizon $r=r_g$. We
introduce for P the co-moving variables ($\xi ^i)_{i=\overline{1,4}}=\left(
R(r,t),\varphi ,\theta ,c\tau \left( r,t\right) \right) ,$where $\tau $ is
the proper time of P, these coordinates forming, according to RTG, another
coordinate system in Minkowski space-time. In the co-moving reference system
$\xi ^i$, the metric tensor $\gamma _{ij}$ of the Minkowski space-time
hasn't the form (\ref{*}) it is determined from the tensor transformation
law. In principle, the one-to-one transformation between the system of
coordinates $x^i$ and $\xi ^i$ can be established using the fact that the
metric coefficients $g_{ij}\left( \xi \right) $ must satisfy the general
covariant Eqs. of RTG which tell us that a gravitational field can have only
spin states 0 and 2, i.e. Eqs. $D_i\tilde{g}^{ij}=0,$ $i,j=\overline{1,4}$.
The connection between the proper time interval $d\tau $ and $dt$ is $d\tau
=\left( 1-\frac{2kM}{c^2}\frac 1r+\frac{kQ^2}{c^4}\frac 1{r^2}\right) dt$.
Thus P can't continue the trajectory beyond the horizon $r=r_g$ because in
this region, the expression $1-\frac{2kM}{c^2}\frac 1r$ $+\frac{kQ^2}{c^4}%
\frac 1{r^2}$ $\,$takes negative values and therefore the proper time $\tau $
can not be a time which P measures on its own clock.

As in Section 3, let us now explore Eq. (\ref{2.14}) with  to
find its solution.

Differentiating this Eq. with respect to $\theta $ and then removing the
common factor $2\frac{dZ}{d\theta }$, we obtain:

\begin{equation}
\hspace{-1.2cm}\frac{d^2Z}{d\theta ^2}+Z=-\frac{2kQ^2}{c^4}Z^3+\frac{3kM}{c^2}Z^2+\frac{Q^2%
}{\mathcal{L}^2c^4m^2}\left( q^2-km^2\right) Z+\frac{kM}{\mathcal{L}^2c^2}-%
\frak{E}\frac{qQ}{m\mathcal{L}^2c^2}.  \label{2.15}
\end{equation}

Let us now find a solution of order $\frac{velocity^{2}}{c^{2}}$ for this
relativistic Eq..

Allowing for the notations (\ref{1.16}), (\ref{2.13}), the expression of $Z$
depending on $u$ is:

\begin{equation}
Z=\frac{u}{1+u\frac{kM}{c^{2}}}  \label{2.16}
\end{equation}

As we have already discussed in the Section 3, we assume that the order of
magnitude of $kMu$ is $v_1^2,$where $v_1$ is a velocity much smaller than
the velocity of light in vacuum. Keeping only to the terms of order $\frac{v_1^2%
}{c^2}$ in (\ref{2.16}), we get for $Z:$

\begin{equation}
Z=u-\frac{kM}{c^{2}}u^{2}.  \label{2.17}
\end{equation}

From (\ref{2.17}), we get to terms of order $\frac{v_1^2}{c^2},$:

\begin{equation}
\frac{d^{2}Z}{d\theta ^{2}}=\left( 1-\frac{2kM}{c^{2}}u\right) \frac{d^{2}u}{%
d\theta ^{2}}-\frac{2kM}{c^{2}}\left( \frac{du}{d\theta }\right) ^{2}.
\label{2.18}
\end{equation}

Thus, by virtue of (\ref{2.17}), (\ref{2.18}), in an approximation of order $%
\frac{velocity^2}{c^2}$ , Eq. (\ref{2.15}) becomes:

\begin{eqnarray}
\left( 1-\frac{2kM}{c^2}u\right) \frac{d^2u}{d\theta ^2}+u&=&\frac{kM}{%
\mathcal{L}^2c^2}-\frak{E}\frac{qQ}{m\mathcal{L}^2c^2}+\frac{Q^2}{\mathcal{L}%
^2c^4m^2}\left( q^2-km^2\right) u+\nonumber\\
&+&4\frac{kM}{c^2}u^2+\frac{2kM}{c^2}\left(
\frac{du}{d\theta }\right) ^2.  \label{2.19}
\end{eqnarray}

In the case of slow motion in weak gravitational fields, Eq.(\ref{2.19})
must reduce to the classical Eq. (\ref{13}). As in Section 3, this happens
for $\frak{E}\simeq 1$ and $\frac 1{\mathcal{L}^2c^2}\simeq \frac 1{J^2}.$
Taking also into account the notations (\ref{1.22}) for the small
dimensionless quantities, let us find an approximate solution to order $%
\frac{velocity^2}{c^2}$ for Eq.:

\begin{equation}
\hspace{-1.2cm}\frac{d^{2}u}{d\theta ^{2}}\left( 1-2\delta \frac{J^{2}}{kM}u\right)
+u\left( 1-\varepsilon \right) =\frac{kM}{J^{2}}-\frac{qQ}{mJ^{2}}+4\delta
\frac{J^{2}}{kM}u^{2}+2\delta \frac{J^{2}}{kM}\left( \frac{du}{d\theta }%
\right) ^{2}.  \label{2.20}
\end{equation}

To solve this we assume a solution of the form:

\begin{equation}
u(\theta )=u_o(\theta )+\varepsilon V(\theta )+\delta (\theta
)+O(\varepsilon ^2)+O(\delta ^2)+O(\varepsilon \delta ).  \label{2.21}
\end{equation}

Substituting this form for $u$ in the differential Eq.(\ref{2.20})and
keeping only the terms of order 0 and 1 in $\varepsilon $ and $\delta $, we
find:

\begin{eqnarray}
&&\frac{d^2u_o}{d\theta ^2}+\varepsilon \frac{d^2V}{d\theta ^2}+\delta \frac{%
d^2\mathcal{W}}{d\theta ^2}-2\delta \frac{J^2}{kM}\frac{d^2u_o}{d\theta ^2}%
u_o+u_o+\varepsilon V+\delta \mathcal{W}-\varepsilon u_o=\nonumber\\
&&=\frac{kM}{J^2}-%
\frac{qQ}{mJ^2}+4\delta \frac{J^2}{kM}u_o^2+2\delta \frac{J^2}{kM}\left(
\frac{du_o}{d\theta }\right) ^2.  \label{2.22}
\end{eqnarray}

Equating the zeroth order terms in $\varepsilon $ and $\delta $ we get Eq. (%
\ref{13}) with the solution (\ref{19}). Equating the first order terms in $%
\varepsilon $ and taking into account (\ref{19}), we get Eq. (\ref{1.28})
with the nonhomogeneous solution (\ref{1.29}), (\ref{1.30}).

Similarly, equating the first order terms in $\delta $ and by virtue of (\ref
{19}), we obtain:

\begin{eqnarray}
\frac{d^2\mathcal{W}}{d\theta ^2}+\mathcal{W}&=&4\frac{J^2}{kM}\left( \frac{kM%
}{J^2}-\frac{qQ}{mJ^2}\right) ^2+2\frac{J^2}{kM}\mathcal{A}^2+\nonumber\\
&&+6\frac{J^2}{kM}%
\left( \frac{kM}{J^2}-\frac{qQ}{mJ^2}\right) \mathcal{A}\cos \theta
\label{2.23}
\end{eqnarray}
with the nonhomogeneous solution:

\begin{equation}
\mathcal{W}(\theta )=\mathcal{W}_1(\theta )+\mathcal{W}_2(\theta ),
\label{2.24}
\end{equation}
where
\begin{eqnarray}
\mathcal{W}_1(\theta ) &=&4\frac{J^2}{kM}\left( \frac{kM}{J^2}-\frac{qQ}{mJ^2%
}\right) ^2+2\frac{J^2}{kM}\mathcal{A}^2  \nonumber \\
\mathcal{W}_2(\theta ) &=&3\frac{J^2}{kM}\left( \frac{kM}{J^2}-\frac{qQ}{mJ^2%
}\right) \mathcal{A}\theta \sin \theta  \label{2.25}
\end{eqnarray}

Introducing (\ref{19}), (\ref{1.29}), (\ref{2.24}) into (\ref{2.21}), we
obtain the solution for the orbit to first order in $\varepsilon $ and $%
\delta $:

\begin{eqnarray}
\hspace{-1cm}u(\theta ) &=&\left[ \frac{kM}{J^{2}}-\frac{qQ}{mJ^{2}}+\varepsilon \left(
\frac{kM}{J^{2}}-\frac{qQ}{mJ^{2}}\right) +\delta \frac{4J^{2}}{kM}\left(
\left( \frac{kM}{J^{2}}-\frac{qQ}{mJ^{2}}\right) ^{2}+\frac{\mathcal{A}^{2}}{%
2}\right) \right] +  \nonumber \\
&&+\mathcal{A}\cos \theta +\left[ \frac{\varepsilon }{2}+\delta \frac{3J^{2}%
}{kM}\left( \frac{kM}{J^{2}}-\frac{qQ}{mJ^{2}}\right) \right] \mathcal{A}%
\theta \sin \theta .  \label{2.26}
\end{eqnarray}

As in the framework of GRT, in the solution (\ref{2.26}) only the last term
is nonperiodic. In fact, this last term is exactly as in (\ref{1.34}).
Therefore, using (\ref{1.35}) the solution (\ref{2.26}) may by written in
the form (\ref{1.36}). The only small difference between GRT's solution and
RTG's solution is in the periodic term of order $\delta .$ But the effect of
these terms is to introduce small periodic variations in the radial distance
of P and it is difficult to be detected. For an orbit of elliptic type,
keeping the terms to first order in $\varepsilon $ and $\delta ,$ the
interval between successive perihelia is given by (\ref{1.38}).

In conclusion, the predictions of RTG in the considered problem are the same
as in GRT in the approximation of order $\frac{velocity^2}{c^2}$ .

\section{Second order approximation of the solution in GRT and RTG}

Let us see if the solutions in GRT and RTG remain the same if we consider them
in an approximation of order $\frac{velocity^4}{c^4}$ .

Let us return to Section 3 at Eq. (\ref{1.21}). If we want to find a
solution of Eq. (\ref{1.21}) to order $\frac{velocity^4}{c^4}$, $velocity\ll
c$, we will not neglect the last term from this Eq. We define the small
dimensionless quantity $\varepsilon $, $\delta $ as in (\ref{1.22}) and in
addition :

\begin{equation}
\zeta =\frac{k^{3}Q^{2}M^{2}}{J^{4}c^{4}}.  \label{3.1}
\end{equation}

Thus Eq. (\ref{1.21}) becomes:

\begin{equation}
\frac{d^{2}u}{d\theta ^{2}}+u\left( 1-\varepsilon \right) =\frac{kM}{J^{2}}-%
\frac{qQ}{mJ^{2}}+3\delta \frac{J^{2}}{kM}u^{2}-2\zeta \left( \frac{J^{2}}{kM%
}\right) ^{2}u^{3}.  \label{3.2}
\end{equation}

To find a solution of this nonlinear Eq. to order $\frac{velocity^{4}}{c^{4}}
$ , we assume a solution of the form:

\begin{eqnarray}
\hspace{-0.7cm}u(\theta ) &=&u_o(\theta )+\varepsilon V(\theta )+\delta W(\theta )+\zeta
\Upsilon (\theta )+\varepsilon ^2S(\theta )+\delta ^2X(\theta )+\varepsilon
\delta Y(\theta )+  \nonumber \\
\hspace{-0.7cm}&&+O(\varepsilon ^3)+O(\delta ^3)+O(\zeta ^2)+O(\zeta \varepsilon )+O(\zeta
\delta )+O(\varepsilon ^2\delta )+O(\varepsilon \delta ^2).  \label{3.3}
\end{eqnarray}

Substituting this form for $u$ in the differential Eq.(\ref{3.2}) and
keeping only the terms to order 2 in $\varepsilon $ and $\delta $ and to
order 1 in $\zeta $, we find:

\begin{eqnarray}
&&\frac{d^2u_o}{d\theta ^2}+\varepsilon \frac{d^2V}{d\theta ^2}+\delta \frac{%
d^2W}{d\theta ^2}+\zeta \frac{d^2\Upsilon }{d\theta ^2}+\varepsilon ^2\frac{%
d^2S}{d\theta ^2}+\delta ^2\frac{d^2X}{d\theta ^2}+\varepsilon \delta \frac{%
d^2Y}{d\theta ^2}+u_o +\nonumber \\
&&+\varepsilon V+\delta W+\zeta \Upsilon +\varepsilon ^2S+\delta
^2X+\varepsilon \delta Y-\varepsilon u_o-\varepsilon ^2V-\varepsilon \delta W
=\frac{kM}{J^2}-\nonumber \\
&&-\frac{qQ}{mJ^2}+3\delta \frac{J^2}{kM}u_o^2+6\varepsilon
\delta \frac{J^2}{kM}u_oV+6\delta ^2\frac{J^2}{kM}u_oW-2\zeta \left( \frac{%
J^2}{kM}\right) ^2u_o^3.  \label{3.4}
\end{eqnarray}

Equating the zeroth order terms in $\varepsilon $, $\delta $ and $\zeta $ we
get Eq. (\ref{13}) with the solution (\ref{19}). Equating the first order
terms in $\varepsilon $ and $\delta ,$ and taking into account (\ref{19}),
we get Eq. (\ref{1.28}), respectively (\ref{1.31}) with the nonhomogeneous
solution (\ref{1.29}), (\ref{1.30}), respectively (\ref{1.32}), (\ref{1.33}%
). Similarly, taking into account (\ref{19}), for the terms of order $\zeta $
we obtain:

\begin{eqnarray}
\frac{d^2\Upsilon }{d\theta ^2}+\Upsilon &=&-2\left( \frac{J^2}{kM}\right)
^2\left[ \left( \frac{kM}{J^2}-\frac{qQ}{mJ^2}\right) ^3+\frac{3\mathcal{A}^2%
}2\left( \frac{kM}{J^2}-\frac{qQ}{mJ^2}\right) \right] -  \nonumber \\
&&-2\left( \frac{J^2}{kM}\right) ^2\left( 3\left( \frac{kM}{J^2}-\frac{qQ}{%
mJ^2}\right) ^2\mathcal{+}\frac{3\mathcal{A}^2}4\right) \mathcal{A}\cos
\theta -  \nonumber \\
&&-3\left( \frac{J^2}{kM}\right) ^2\left( \frac{kM}{J^2}-\frac{qQ}{mJ^2}%
\right) \mathcal{A}^2\cos 2\theta +  \nonumber \\
&&-\frac 12\left( \frac{J^2}{kM}\right) ^2\mathcal{A}^3\cos 3\theta
\label{3.5}
\end{eqnarray}
with the nonhomogeneous solution :

\begin{equation}
\Upsilon (\theta )=\Upsilon _1(\theta )+\Upsilon _2(\theta )+\Upsilon
_3(\theta )+\Upsilon _4(\theta ),  \label{3.6}
\end{equation}
where

\begin{eqnarray}
\Upsilon _1(\theta ) &=&-2\left( \frac{J^2}{kM}\right) ^2\left[ \left( \frac{%
kM}{J^2}-\frac{qQ}{mJ^2}\right) ^3+\frac{3\mathcal{A}^2}2\left( \frac{kM}{J^2%
}-\frac{qQ}{mJ^2}\right) \right]  \nonumber \\
\Upsilon _2(\theta ) &=&-\left( \frac{J^2}{kM}\right) ^2\left( 3\left( \frac{%
kM}{J^2}-\frac{qQ}{mJ^2}\right) ^2\mathcal{+}\frac{3\mathcal{A}^2}4\right)
\mathcal{A}\theta \sin \theta  \nonumber \\
\Upsilon _3(\theta ) &=&\left( \frac{J^2}{kM}\right) ^2\left( \frac{kM}{J^2}-%
\frac{qQ}{mJ^2}\right) \mathcal{A}^2\cos 2\theta  \nonumber \\
\Upsilon _4(\theta ) &=&\frac 1{16}\left( \frac{J^2}{kM}\right) ^2\mathcal{A}%
^3\cos 3\theta  \label{3.7}
\end{eqnarray}

Identifying the second order terms in $\varepsilon ^2$ and allowing for (\ref
{1.29}), (\ref{1.30}), it results Eq.:

\begin{equation}
\frac{d^2S}{d\theta ^2}+S=\frac{kM}{J^2}-\frac{qQ}{mJ^2}+\frac{\mathcal{A}}%
2\theta \sin \theta  \label{3.8}
\end{equation}
with the nonhomogeneous solution :

\begin{equation}
S(\theta )=S_1(\theta )+S_2(\theta ),  \label{3.9}
\end{equation}
where

\begin{eqnarray}
S_1(\theta ) &=&\frac{kM}{J^2}-\frac{qQ}{mJ^2}  \nonumber \\
S_2(\theta ) &=&\frac{\mathcal{A}}8\left( \theta \sin \theta -\theta ^2\cos
\theta \right).  \label{3.10}
\end{eqnarray}

Equating the second order terms in $\delta ^2$, by virtue of (\ref{19}) and (%
\ref{1.32}), (\ref{1.33}), we get:

\begin{eqnarray}
\frac{d^2X}{d\theta ^2}+X &=&18\left( \frac{J^2}{kM}\right) ^2\left( \frac{kM%
}{J^2}-\frac{qQ}{mJ^2}\right) \left[ \left( \frac{kM}{J^2}-\frac{qQ}{mJ^2}%
\right) ^2+\frac{\mathcal{A}^2}2\right] +  \nonumber \\
&&+18\left( \frac{J^2}{kM}\right) ^2\left( \frac{kM}{J^2}-\frac{qQ}{mJ^2}%
\right) ^2\mathcal{A}\theta \sin \theta +  \nonumber \\
&&+18\left( \frac{J^2}{kM}\right) ^2\left( \left( \frac{kM}{J^2}-\frac{qQ}{%
mJ^2}\right) ^2+\frac 5{12}\mathcal{A}^2\right) \mathcal{A}\cos \theta +
\nonumber \\
&&+9\left( \frac{J^2}{kM}\right) ^2\left( \frac{kM}{J^2}-\frac{qQ}{mJ^2}%
\right) \mathcal{A}^2\theta \sin 2\theta -  \nonumber \\
&&-3\left( \frac{J^2}{kM}\right) ^2\left( \frac{kM}{J^2}-\frac{qQ}{mJ^2}%
\right) \mathcal{A}^2\cos 2\theta -  \nonumber \\
&&-\frac 32\left( \frac{J^2}{kM}\right) ^2\mathcal{A}^3\cos 3\theta .
\label{3.11}
\end{eqnarray}

The nonhomogeneous solution of (\ref{3.11}) can be easily checked:

\begin{equation}
X(\theta )=X_1(\theta )+X_2(\theta )+X_3(\theta )+X_4(\theta )+X_5(\theta
)+X_6(\theta ),  \label{3.12}
\end{equation}
where
\begin{eqnarray}
X_1(\theta ) &=&18\left( \frac{J^2}{kM}\right) ^2\left( \frac{kM}{J^2}-\frac{%
qQ}{mJ^2}\right) \left[ \left( \frac{kM}{J^2}-\frac{qQ}{mJ^2}\right) ^2+%
\frac{\mathcal{A}^2}2\right]  \nonumber \\
X_2(\theta ) &=&\frac 92\left( \frac{J^2}{kM}\right) ^2\left( \frac{kM}{J^2}-%
\frac{qQ}{mJ^2}\right) ^2\mathcal{A}\left( \theta \sin \theta -\theta ^2\cos
\theta \right)  \nonumber \\
X_3(\theta ) &=&9\left( \frac{J^2}{kM}\right) ^2\left( \left( \frac{kM}{J^2}-%
\frac{qQ}{mJ^2}\right) ^2+\frac 5{12}\mathcal{A}^2\right) \mathcal{A}\theta
\sin \theta  \nonumber \\
X_4(\theta ) &=&\left( \frac{J^2}{kM}\right) ^2\left( \frac{kM}{J^2}-\frac{qQ%
}{mJ^2}\right) \mathcal{A}^2\left( -3\theta \sin 2\theta -4\cos 2\theta
\right)  \nonumber \\
X_5(\theta ) &=&\left( \frac{J^2}{kM}\right) ^2\left( \frac{kM}{J^2}-\frac{qQ%
}{mJ^2}\right) \mathcal{A}^2\cos 2\theta  \nonumber \\
X_6(\theta ) &=&\frac 3{16}\left( \frac{J^2}{kM}\right) ^2\mathcal{A}^3\cos
3\theta  \label{3.13}
\end{eqnarray}

Equating the terms in $\varepsilon \delta $ and by virtue of (\ref{19}), (%
\ref{1.29}), (\ref{1.30}), (\ref{1.32}), (\ref{1.33}), we get:

\begin{eqnarray}
\frac{d^2Y}{d\theta ^2}+Y &=&\frac{J^2}{kM}\left[ 9\left( \frac{kM}{J^2}-%
\frac{qQ}{mJ^2}\right) ^2+\frac 32\mathcal{A}^2\right] +  \nonumber \\
&&+6\frac{J^2}{kM}\left( \frac{kM}{J^2}-\frac{qQ}{mJ^2}\right) \mathcal{A}%
\theta \sin \theta -  \nonumber \\
&&-\frac 12\frac{J^2}{kM}\mathcal{A}^2\cos 2\theta +  \nonumber \\
&&+6\frac{J^2}{kM}\left( \frac{kM}{J^2}-\frac{qQ}{mJ^2}\right) \mathcal{A}%
\cos \theta +  \nonumber \\
&&+\frac 32\frac{J^2}{kM}\mathcal{A}^2\theta \sin 2\theta .  \label{3.14}
\end{eqnarray}
having the following nonhomogeneous solution:

\begin{equation}
Y(\theta )=Y_1(\theta )+Y_2(\theta )+Y_3(\theta )+Y_4(\theta )+Y_5(\theta )
\label{3.15}
\end{equation}
where
\begin{eqnarray}
Y_1(\theta ) &=&\frac{J^2}{kM}\left[ 9\left( \frac{kM}{J^2}-\frac{qQ}{mJ^2}%
\right) ^2+\frac 32\mathcal{A}^2\right]  \nonumber \\
Y_2(\theta ) &=&\frac 32\frac{J^2}{kM}\left( \frac{kM}{J^2}-\frac{qQ}{mJ^2}%
\right) \mathcal{A}\left( \theta \sin \theta -\theta ^2\cos \theta \right)
\nonumber \\
Y_3(\theta ) &=&\frac 16\frac{J^2}{kM}\mathcal{A}^2\cos 2\theta  \nonumber \\
Y_4(\theta ) &=&3\frac{J^2}{kM}\left( \frac{kM}{J^2}-\frac{qQ}{mJ^2}\right)
\mathcal{A}\cos \theta  \nonumber \\
Y_5(\theta ) &=&\frac{J^2}{kM}\mathcal{A}^2\left( -\frac 12\theta \sin
2\theta -\frac 23\cos 2\theta \right) .  \label{3.16}
\end{eqnarray}

Introducing (\ref{19}), (\ref{1.29}), (\ref{1.32}), (\ref{3.6}), (\ref{3.9}%
), (\ref{3.12}), (\ref{3.15}) into (\ref{3.4}), we find the nonperiodic term
of the solution:\newline

\begin{eqnarray}
\hspace{-1.4cm}\left[ \frac \varepsilon 2\right.&+&\left.\delta \frac{3J^2}{kM}\left( \frac{kM}{J^2}-%
\frac{qQ}{mJ^2}\right) -\zeta \left( \frac{J^2}{kM}\right) ^2\left( 3\left(
\frac{kM}{J^2}-\frac{qQ}{mJ^2}\right) ^2\mathcal{+}\frac{3\mathcal{A}^2}%
4\right) \right] \mathcal{A}\theta \sin \theta +  \nonumber \\
\hspace{-1.2cm}+\left[ \frac{\varepsilon ^2}8\right.&+&\left.\delta ^2\left( \frac{J^2}{kM}\right)
^2\left( \frac{27}2\left( \frac{kM}{J^2}-\frac{qQ}{mJ^2}\right) ^2+\frac{15}4%
\mathcal{A}^2\right) +\varepsilon \delta \frac 32\frac{J^2}{kM}\left( \frac{%
kM}{J^2}-\frac{qQ}{mJ^2}\right) \right] \mathcal{A}\theta \sin \theta -
\nonumber \\
\hspace{-1cm}-\left[ \frac {\varepsilon^2}{8}\right.&+&\left.\delta ^2\frac 92\left( \frac{J^2}{kM}%
\right) ^2\left( \frac{kM}{J^2}-\frac{qQ}{mJ^2}\right) ^2+\varepsilon \delta
\frac 32\frac{J^2}{kM}\left( \frac{kM}{J^2}-\frac{qQ}{mJ^2}\right) \right]
\mathcal{A}\theta ^2\cos \theta -  \nonumber \\
\hspace{-1.2cm}&&-\left[ 3\delta ^2\left( \frac{J^2}{kM}\right) ^2\left( \frac{kM}{J^2}-%
\frac{qQ}{mJ^2}\right) +\varepsilon \delta \frac 12\frac{J^2}{kM}\right]
\mathcal{A}^2\theta \sin 2\theta .  \label{3.17}
\end{eqnarray}

We note that to the second order in $\varepsilon $ and $\delta ,$ and to the first
order in $\zeta ,$ the following expression$:$

\begin{eqnarray}
&&\cos \left(\theta -\theta \left[\frac \varepsilon 2+\delta \frac{3J^2}{kM}\left(
\frac{kM}{J^2}-\frac{qQ}{mJ^2}\right) -\zeta \left( \frac{J^2}{kM}\right)
^2\left( 3\left( \frac{kM}{J^2}-\frac{qQ}{mJ^2}\right) ^2\right.\right.\right. +  \nonumber \\
\hspace{-1cm}&&+\left.\frac{3%
\mathcal{A}^2}4\right)+\frac{\varepsilon ^2}8+\delta ^2\left( \frac{J^2%
}{kM}\right) ^2\left( \frac{27}2\left( \frac{kM}{J^2}-\frac{qQ}{mJ^2}\right)
^2+\frac{15}4\mathcal{A}^2\right) +\varepsilon \delta \frac 32\frac{J^2}{kM}%
\left( \frac{kM}{J^2}\right.  \nonumber \\
&&\left.\left.\left.-\frac{qQ}{mJ^2}\right) -\left( 6\delta ^2\left( \frac{J^2}{kM}\right)
^2\left( \frac{kM}{J^2}-\frac{qQ}{mJ^2}\right) +\varepsilon \delta \frac{J^2%
}{kM}\right) \mathcal{A}\cos \theta \right]\mbox{\rm   }\right)\mbox{\rm  ,}
\label{3.18}
\end{eqnarray}
reduces to $\cos \theta $ plus the expression written in (\ref{3.17}).
Therefore, the solution is an approximation of order $\frac{velocity^4}{c^4}$
for the \textit{\ electrogravitational Kepler problem }in the framework of
GRT and may be written as:

\begin{eqnarray}
u(\theta ) &=&\frac{kM}{J^2}-\frac{qQ}{mJ^2}+\nonumber\\
&&+\mathcal{A}\cos \left(\theta -\theta
\left[\frac \varepsilon 2+\delta \frac{3J^2}{kM}\left( \frac{kM}{J^2}-\frac{qQ}{%
mJ^2}\right) -\right.\right.  \nonumber \\
&&-\zeta \left( \frac{%
J^2}{kM}\right) ^2\left( 3\left( \frac{kM}{J^2}-\frac{qQ}{mJ^2}\right) ^2%
\mathcal{+}\frac{3\mathcal{A}^2}4\right) +\frac{\varepsilon ^2}8+  \nonumber
\\
&&+\delta ^2\left(
\frac{J^2}{kM}\right) ^2\left( \frac{27}2\left( \frac{kM}{J^2}-\frac{qQ}{mJ^2%
}\right) ^2+\frac{15}4\mathcal{A}^2\right) +  \nonumber \\
&&+\varepsilon \delta
\frac 32\frac{J^2}{kM}\left( \frac{kM}{J^2}-\frac{qQ}{mJ^2}\right) -
\nonumber \\
&&-\left.\left.\left( 6\delta
^2\left( \frac{J^2}{kM}\right) ^2\left( \frac{kM}{J^2}-\frac{qQ}{mJ^2}%
\right) +\varepsilon \delta \frac{J^2}{kM}\right) \mathcal{A}\cos \theta \right]\right)+
\nonumber \\
&&+(%
\mbox{\rm periodic terms of
order }\varepsilon ,\delta ,\varepsilon ^2,\delta ^2,\zeta ) \label{3.19}
\end{eqnarray}

\smallskip Let us now see how   this approximation of order $\frac{%
velocity^4}{c^4}$ for the\textit{\ electrogravitational Kepler problem } looks in
the framework of RTG.

We return to Eq. (\ref{2.15}). Keeping in (\ref{2.16}) to terms of order $%
\frac{velocity^4}{c^4},$ we get for $Z:$

\begin{equation}
Z=u-\frac{kM}{c^{2}}u^{2}+\frac{k^{2}M^{2}}{c^{4}}u^{3}.  \label{3.20}
\end{equation}

From (\ref{3.20}) we get to terms of order $\frac{velocity^4}{c^4}$:

\begin{equation}
\frac{d^{2}Z}{d\theta ^{2}}=\left( 1-\frac{2kM}{c^{2}}u+\frac{3k^{2}M^{2}}{%
c^{4}}u^{2}\right) \frac{d^{2}u}{d\theta ^{2}}+\left( -\frac{2kM}{c^{2}}+%
\frac{6k^{2}M^{2}}{c^{4}}u\right) \left( \frac{du}{d\theta }\right) ^{2}.
\label{3.21}
\end{equation}

Thus, by virtue of (\ref{3.20}), (\ref{3.21}), for $\frak{E}\simeq 1$ and $%
\frac 1{\mathcal{L}^2c^2}\simeq \frac 1{J^2}$, Eq. (\ref{2.15}) becomes, in
an approximation of order $\frac{velocity4}{c^4}$ :

\begin{eqnarray}
&&\frac{d^{2}u}{d\theta ^{2}}\left( 1-2\delta \frac{J^{2}}{kM}u+3\delta
^{2}\left( \frac{J^{2}}{kM}\right) ^{2}u^{2}\right) +u\left( 1-\varepsilon
\right) =  \nonumber \\
&&=\frac{kM}{J^{2}}-\frac{qQ}{mJ^{2}}+4\delta \frac{J^{2}}{kM}%
u^{2}-\varepsilon \delta \frac{J^{2}}{kM}u^{2}-7\delta ^{2}\left( \frac{J^{2}}{kM}\right) ^{2}u^{3}- \nonumber \\
&&\ -2\zeta \left( \frac{%
J^{2}}{kM}\right) ^{2}u^{3}+\left( 2\delta -6\delta ^{2}\frac{J^{2}}{kM}u\right) \frac{J^{2}}{kM}%
\left( \frac{du}{d\theta }\right) ^{2}  \label{3.22}
\end{eqnarray}
 To find a solution of this nonlinear Eq. to order $\frac{%
velocity^{4}}{c^{4}}$ , we assume a solution of the form:

\begin{eqnarray}
u(\theta ) &=&u_o(\theta )+\varepsilon V(\theta )+\delta \mathcal{W}(\theta
)+\zeta \Upsilon (\theta )+\varepsilon ^2S(\theta )+\delta ^2\mathcal{X}%
(\theta )+\varepsilon \delta \mathcal{Y}(\theta )+  \nonumber \\
&&O(\varepsilon ^3)+O(\delta ^3)+O(\zeta ^2)+O(\zeta \varepsilon )+O(\zeta
\delta )+O(\varepsilon ^2\delta )+O(\varepsilon \delta ^2).  \label{3.23}
\end{eqnarray}

Substituting this form for $u$ in the differential Eq.(\ref{3.22}) and
keeping only the terms to the  order 2 in $\varepsilon $ and $\delta $ and to
the order 1 in $\zeta $, we find:

\begin{eqnarray}
&\frac{d^2u_o}{d\theta ^2}+\varepsilon \frac{d^2V}{d\theta ^2}+\delta \frac{%
d^2\mathcal{W}}{d\theta ^2}+\zeta \frac{d^2\Upsilon }{d\theta ^2}%
+\varepsilon ^2\frac{d^2S}{d\theta ^2}+\delta ^2\frac{d^2\mathcal{X}}{%
d\theta ^2}+\varepsilon \delta \frac{d^2\mathcal{Y}}{d\theta ^2}-  \nonumber \\
&-\delta \frac{2J^2}{kM}u_o\frac{d^2u_o}{d\theta ^2}-\varepsilon \delta \frac{2J^2}{kM}V\frac{d^2u_o}{d\theta ^2}-\delta ^2%
\frac{2J^2}{kM}\mathcal{W}\frac{d^2u_o}{d\theta ^2}-\varepsilon \delta \frac{%
2J^2}{kM}u_o\frac{d^2V}{d\theta ^2}- \nonumber \\
&-\delta ^2\frac{2J^2}{kM}u_o\frac{d^2%
\mathcal{W}}{d\theta ^2} +3\delta ^2\left( \frac{J^2}{kM}\right) ^2u_o^2\frac{d^2u_o}{d\theta ^2}%
+u_o+\varepsilon V+\delta \mathcal{W}+\zeta \Upsilon+  \nonumber \\
&+\varepsilon ^2S+\delta
^2\mathcal{X}+\varepsilon \delta \mathcal{Y}-\varepsilon u_o-\varepsilon
^2V-\varepsilon \delta \mathcal{W}= \nonumber\\
&=\frac{kM}{J^2}-\frac{qQ}{mJ^2}+\delta \frac{4J^2}{kM}u_o^2+\varepsilon
\delta \frac{8J^2}{kM}u_oV+\delta ^2\frac{8J^2}{kM}u_o\mathcal{W}%
-\varepsilon \delta \frac{J^2}{kM}u_o^2-  \nonumber \\
&-7\delta ^2\left( \frac{J^2}{kM}%
\right) ^2u_o^3-2\zeta \left( \frac{J^2}{kM}\right) ^2u_o^3+\delta \frac{2J^2}{kM}\left(
\frac{du_o}{d\theta }\right) ^2+\varepsilon \delta \frac{4J^2}{kM}\frac{du_o%
}{d\theta }\frac{dV}{d\theta }+  \nonumber \\
&+\delta ^2\frac{4J^2}{kM}\frac{du_o}{d\theta }%
\frac{d\mathcal{W}}{d\theta }-6\delta ^2\left( \frac{J^2}{kM}\right) ^2u_o\left( \frac{du_o}{d\theta }%
\right) ^2.  \label{3.24}
\end{eqnarray}

Equating the zeroth order terms in $\varepsilon $, $\delta $ and $\zeta $ we
get Eq. (\ref{13}) with the solution (\ref{19}). Equating the first order
terms in $\varepsilon $ and $\delta ,$ and taking into account (\ref{19}),
we get Eq. (\ref{1.28}), respectively (\ref{2.23}) with the nonhomogeneous
solution (\ref{1.29}), (\ref{1.30}), respectively (\ref{2.24}), (\ref{2.25}%
). For the terms of order $\zeta $ we obtain (\ref{3.5}), with the
nonhomogeneous solution (\ref{3.6}), (\ref{3.7}). Identifying the second
order terms in $\varepsilon ^2$ and allowing for (\ref{1.29}), (\ref{1.30}),
it results Eq. (\ref{3.8}), having the nonhomogeneous solution (\ref{3.9}), (%
\ref{3.10}).

Equating the second order terms in $\delta ^{2}$, we get:

\begin{eqnarray}
\frac{d^2\mathcal{X}}{d\theta ^2}+\mathcal{X} &=&\left( \frac{J^2}{kM}%
\right) ^2\left( \frac{kM}{J^2}-\frac{qQ}{mJ^2}\right) \left[ 25\left( \frac{%
kM}{J^2}-\frac{qQ}{mJ^2}\right) ^2+\frac{11}2\mathcal{A}^2\right] +
\nonumber \\
&&+18\left( \frac{J^2}{kM}\right) ^2\left( \frac{kM}{J^2}-\frac{qQ}{mJ^2}%
\right) ^2\mathcal{A}\theta \sin \theta +  \nonumber \\
&&+\left( \frac{J^2}{kM}\right) ^2\left( 18\left( \frac{kM}{J^2}-\frac{qQ}{%
mJ^2}\right) ^2+\frac{15}2\mathcal{A}^2\right) \mathcal{A}\cos \theta +
\nonumber \\
&&+\frac{15}2\left( \frac{J^2}{kM}\right) ^2\left( \frac{kM}{J^2}-\frac{qQ}{%
mJ^2}\right) \mathcal{A}^2\cos 2\theta -  \nonumber \\
&&+\frac 12\left( \frac{J^2}{kM}\right) ^2\mathcal{A}^3\cos 3\theta .
\label{3.25}
\end{eqnarray}

The nonhomogeneous solution of (\ref{3.25}) can be easily checked:

\begin{equation}
\mathcal{X}(\theta )=\mathcal{X}_1(\theta )+\mathcal{X}_2(\theta )+\mathcal{X%
}_3(\theta )+\mathcal{X}_4(\theta )+\mathcal{X}_5(\theta ),  \label{3.26}
\end{equation}
where

\begin{eqnarray}
\mathcal{X}_1(\theta ) &=&\left( \frac{J^2}{kM}\right) ^2\left( \frac{kM}{J^2%
}-\frac{qQ}{mJ^2}\right) \left[ 25\left( \frac{kM}{J^2}-\frac{qQ}{mJ^2}%
\right) ^2+\frac{11}2\mathcal{A}^2\right]  \nonumber \\
\mathcal{X}_2(\theta ) &=&\frac 92\left( \frac{J^2}{kM}\right) ^2\left(
\frac{kM}{J^2}-\frac{qQ}{mJ^2}\right) ^2\mathcal{A}\left( \theta \sin \theta
-\theta ^2\cos \theta \right)  \nonumber \\
\mathcal{X}_3(\theta ) &=&\left( \frac{J^2}{kM}\right) ^2\left( 9\left(
\frac{kM}{J^2}-\frac{qQ}{mJ^2}\right) ^2+\frac{15}4\mathcal{A}^2\right)
\mathcal{A}\theta \sin \theta  \nonumber \\
\mathcal{X}_4(\theta ) &=&-\frac 52\left( \frac{J^2}{kM}\right) ^2\left(
\frac{kM}{J^2}-\frac{qQ}{mJ^2}\right) \mathcal{A}^2\cos 2\theta  \nonumber \\
\mathcal{X}_5(\theta ) &=&-\frac 1{16}\left( \frac{J^2}{kM}\right) ^2%
\mathcal{A}^3\cos 3\theta .  \label{3.27}
\end{eqnarray}

Equating the terms in $\varepsilon \delta $ and by virtue of (\ref{19}), (%
\ref{1.29}), (\ref{1.30}), (\ref{2.24}), (\ref{2.25}), we get:

\begin{eqnarray}
\frac{d^2\mathcal{Y}}{d\theta ^2}+\mathcal{Y} &=&\frac{J^2}{kM}\left[
11\left( \frac{kM}{J^2}-\frac{qQ}{mJ^2}\right) ^2+\frac 32\mathcal{A}%
^2\right] +  \nonumber \\
&&+6\frac{J^2}{kM}\left( \frac{kM}{J^2}-\frac{qQ}{mJ^2}\right) \mathcal{A}%
\theta \sin \theta +  \nonumber \\
&&+\frac{J^2}{kM}\mathcal{A}^2\cos 2\theta +  \nonumber \\
&&+6\frac{J^2}{kM}\left( \frac{kM}{J^2}-\frac{qQ}{mJ^2}\right) \mathcal{A}%
\cos \theta +  \nonumber \\
&&+\frac 12\frac{J^2}{kM}\mathcal{A}^2\theta \sin 2\theta .  \label{3.28}
\end{eqnarray}
with the following nonhomogeneous solution:

\begin{equation}
\mathcal{Y}(\theta )=\mathcal{Y}_1(\theta )+\mathcal{Y}_2(\theta )+\mathcal{Y%
}_3(\theta )+\mathcal{Y}_4(\theta )+\mathcal{Y}_5(\theta )  \label{3.29}
\end{equation}
where

\begin{eqnarray}
\mathcal{Y}_1(\theta ) &=&\frac{J^2}{kM}\left[ 11\left( \frac{kM}{J^2}-\frac{%
qQ}{mJ^2}\right) ^2+\frac 32\mathcal{A}^2\right]  \nonumber \\
\mathcal{Y}_2(\theta ) &=&\frac 32\frac{J^2}{kM}\left( \frac{kM}{J^2}-\frac{%
qQ}{mJ^2}\right) \mathcal{A}\left( \theta \sin \theta -\theta ^2\cos \theta
\right)  \nonumber \\
\mathcal{Y}_3(\theta ) &=&-\frac 13\frac{J^2}{kM}\mathcal{A}^2\cos 2\theta
\nonumber \\
\mathcal{Y}_4(\theta ) &=&3\frac{J^2}{kM}\left( \frac{kM}{J^2}-\frac{qQ}{mJ^2%
}\right) \mathcal{A}\cos \theta  \nonumber \\
\mathcal{Y}_5(\theta ) &=&\frac{J^2}{kM}\mathcal{A}^2\left( -\frac 16\theta
\sin 2\theta -\frac 29\cos 2\theta \right) .  \label{3.30}
\end{eqnarray}

Introducing (\ref{19}), (\ref{1.29}), (\ref{2.24}), (\ref{3.6}), (\ref{3.9}%
), (\ref{3.26}), (\ref{3.29}) into (\ref{3.23}), we find the nonperiodic
term of this solution:

\begin{eqnarray}
&&\left[ \frac \varepsilon 2+\delta \frac{3J^2}{kM}\left( \frac{kM}{J^2}-%
\frac{qQ}{mJ^2}\right) -\zeta \left( \frac{J^2}{kM}\right) ^2\left( 3\left(
\frac{kM}{J^2}-\frac{qQ}{mJ^2}\right) ^2+ \right.\right. \nonumber \\
&&\left.\left.+\frac{3\mathcal{A}^2}%
4\right) \right] \mathcal{A}\theta \sin \theta +\left[ \frac{\varepsilon ^2}8+\delta ^2\left( \frac{J^2}{kM}\right)
^2\left( \frac{27}2\left( \frac{kM}{J^2}-\frac{qQ}{mJ^2}\right) ^2+\frac{15}4%
\mathcal{A}^2\right)\right. +
\nonumber \\
&&\left.+\varepsilon \delta \frac 32\frac{J^2}{kM}\left( \frac{%
kM}{J^2}-\frac{qQ}{mJ^2}\right) \right] \mathcal{A}\theta \sin \theta -\left[ \varepsilon ^2\frac 18+\delta ^2\frac 92\left( \frac{J^2}{kM}%
\right) ^2\left( \frac{kM}{J^2}-\frac{qQ}{mJ^2}\right) ^2+\right. \nonumber \\
&&\left.+\varepsilon \delta
\frac 32\frac{J^2}{kM}\left( \frac{kM}{J^2}-\frac{qQ}{mJ^2}\right) \right]
\mathcal{A}\theta ^2\cos \theta -\varepsilon \delta \frac 16\frac{J^2}{kM}\mathcal{A}^2\theta \sin 2\theta
.  \label{3.31}
\end{eqnarray}

We note that to the second order in $\varepsilon $ and $\delta ,$ and to the first
order in $\zeta ,$ the following expression$:$

\begin{eqnarray}
&&\cos \left(\theta -\theta \left[\frac \varepsilon 2+\delta \frac{3J^2}{kM}\left(
\frac{kM}{J^2}-\frac{qQ}{mJ^2}\right) -\zeta \left( \frac{J^2}{kM}\right)
^2\left( 3\left( \frac{kM}{J^2}-\frac{qQ}{mJ^2}\right) ^2+\right.\right.\right. \nonumber \\
&&\left.+\frac{3%
\mathcal{A}^2}4\right) +\frac{\varepsilon ^2}8+\delta ^2\left( \frac{J^2%
}{kM}\right) ^2\left( \frac{27}2\left( \frac{kM}{J^2}-\frac{qQ}{mJ^2}\right)
^2+\frac{15}4\mathcal{A}^2\right) + \nonumber \\
&&+\left.\left.\varepsilon \delta \frac 32\frac{J^2}{kM}%
\left( \frac{kM}{J^2}-\frac{qQ}{mJ^2}\right)-\varepsilon \delta \frac 13\frac{J^2}{kM}%
\mathcal{A}\cos \mathcal{\theta }\right]\right)\mbox{\rm  ,}  \label{3.32}
\end{eqnarray}
reduces to $\cos \theta $ plus the expression written in (\ref{3.31}).
Therefore, the solution in an approximation of order $\frac{velocity^4}{c^4}$
for the \textit{\ electrogravitational Kepler problem } and in the framework of
RTG may be written as:

\begin{eqnarray}
u(\theta )&=&\frac{kM}{J^2}-\frac{qQ}{mJ^2}+\nonumber\\
&&+\mathcal{A}\cos\left(\theta -\theta
\left[\frac \varepsilon 2+\delta \frac{3J^2}{kM}\left( \frac{kM}{J^2}-\frac{qQ}{%
mJ^2}\right) -\right.\right.  \nonumber \\
&&-\zeta \left( \frac{%
J^2}{kM}\right) ^2\left( 3\left( \frac{kM}{J^2}-\frac{qQ}{mJ^2}\right) ^2%
\mathcal{+}\frac{3\mathcal{A}^2}4\right) +\frac{\varepsilon ^2}8+  \nonumber
\\
&&+\delta ^2\left(
\frac{J^2}{kM}\right) ^2\left( \frac{27}2\left( \frac{kM}{J^2}-\frac{qQ}{mJ^2%
}\right) ^2+\frac{15}4\mathcal{A}^2\right) +  \nonumber \\
&&\left.\left.+\varepsilon \delta
\frac 32\frac{J^2}{kM}\left( \frac{kM}{J^2}-\frac{qQ}{mJ^2}\right)
-\varepsilon \delta \frac 13\frac{J^2}{kM}\mathcal{A}\cos \mathcal{\theta }%
\right]\right)+  \nonumber \\
&&+(\mbox{\rm periodic terms of
order }\varepsilon ,\delta ,\varepsilon ^2,\delta ^2,\zeta )  \label{3.33}
\end{eqnarray}

This solution must be compared with the solution (\ref{3.19}) obtained in
the framework of GRT. As we can see, considering approximate  solutions to
order $\frac{velocity^4}{c^4},$ $velocity\ll c,$ the shift of the perihelion
has different values in the two theories.

\smallskip

\section{Conclusions}

\smallskip

We can conclude that the orbits described by the charged mass point with
mass $m$ and electric charge $q$, in the electrogravitational field produced
by the charged mass point with $M$ and electric charge $Q$, have the same
shape in GRT and RTG. In RTG, the range of variable $r$ is moved back, with
respect to the one in GRT, with $\frac{kM}{c^2}$.

The substantial difference between the solution in GRT and RTG is
established in the region close to the gravitational radius, $r_g$. From the
viewpoint of GRT, P crosses the horizon $r=r_g$ only in the inside
direction. This trajectory doesn't reach the singularity $r=0$; it ends at
some point inside the region with $r$ in the interval $\left( 0,\frac{kM}{c^2%
}-\frac 1{c^2}\sqrt{k^2M^2-kQ^2}\right) .$ From the viewpoint of RTG, the
trajectory of P can't continue beyond the horizon $r=r_g.$

The orbits of elliptic type of P, rotate slowly in the same direction  or in the
opposite directions  in which they are described.

In an approximation of the solution for the \textit{electrogravitational
Kepler problem }to the first order, the advance of perihelion per revolution is
the same in GRT and RTG. In an approximation of the solution to the second order
this advance of perihelion differs in the two theories.

\smallskip

\textbf{Acknowledgements. }I would like to thank my teacher E. So\'{o}s, who
has directed and guided me in this field, for helpful discussions and
critical remarks which made me undertake and finish this work.

\smallskip

\end{document}